------------------------------------------------------------------------------

\documentstyle[12pt]{article}
\begin{document}
\title{\bf COMPACT OBJECTS  WITH LARGE MASSES. }
\author{L.V.Verozub}
\maketitle
\centerline{\em Department of Physics and Astronomy Kharkov State University}
{\em  Kharkov  310077  Ukraine}
\begin{abstract}

{\em  It is shown that gravitational theory allows  stable equilibrium
configurations of the degenerated Fermi- gas, whose masses $M \gg  M_{\odot}$
and sizes are less than  $\alpha  = 2GM/c$  ( $G$ is the gravitational
constant
and $c$ is the speed of light). They have no events horizon at the distance
$ r = \alpha $  from the center. This are objects with low luminosity and
are a new candidate in Dark Matter im the Universe.
Orbits of test particles near the above  objects are considered.}
\end{abstract}
\section{ Introduction.}
     Our knowledge about the limiting masses of equilibrium stars configu-
rations and about the fate of collapsing objects is based on the predict-
ions of Einstein's general relativity. The astrophysical data   do  not
contradict   these results, however, they do not prove their validity
unambiguously. When analyzing the observation data it is necessary to take
into consideration the fact that there also exist other gravitation theor-
ies which satisfy (like the general relativity) the post-Newtonian tests,
predicting, however, other physical results in strong fields.(See, for
example , \cite{Rozen}  and \cite{Chang} ). In this  paper,
starting from the gravitational equations proposed before \cite{Verozub} ,
we show that stable equilibrium configurations of a degenerated Fermi-gas
can exist with the following properties: their masses $M$ are many times
as large as the  Sun and their sizes are smaller  than
$\alpha  = 2GM/c$ ,
where $G$ is the gravitational constant and c is the speed of light.

\section{Spherically-Symmetric Gravitational Field.}

     The  above mentioned  gravitation equations are
\begin{equation}
     B^{\gamma }_{\alpha \beta ,\gamma } - B^{\epsilon }_{\alpha \delta }
     B^{\delta }_{\beta \epsilon } = 0
\label{myeqs}
\end{equation}

These are vacuum bimetric equations for the tensor
\begin{equation}
      B^{\gamma }_{\alpha \beta } = \Pi ^{\gamma }_{\alpha \beta } -
       \Pi ^{\gamma }_{\alpha \beta }
\label{tensB}
\end{equation}
(Greek indices run from 0  to 3) , where
\begin{equation}
   \Pi ^{\gamma }_{\alpha \beta } =
\Gamma ^{\gamma }_{\alpha \beta } -
(n+1)^{-1} \left [
\delta ^{\gamma }_{\alpha } \Gamma ^{\epsilon }_{\epsilon \beta } -
\delta ^{\gamma }_{\beta } \Gamma ^{\epsilon }_{\epsilon \alpha }] \right ] ,
\label{Thomases}
\end{equation}
\begin{equation}
\stackrel{\circ}{\Pi} ^{\gamma }_{\alpha \beta } =
\stackrel{\circ}{\Gamma} ^{\gamma }_{\alpha \beta } -
(n+1)^{-1}
\left [ \delta ^{\gamma }_{\alpha } \stackrel{\circ}{\Gamma }^{\epsilon }_
{\epsilon \beta } -
\delta _{\beta }^{\gamma } \stackrel{\circ}{\Gamma }_{\epsilon \alpha }
^{\epsilon }  \right ] ,
\label{Thomases0}
\end{equation}
$\stackrel{\circ}{\Gamma }_{\alpha \beta }^{\gamma }$
 are the Christoffel symbols of the pseudo-Euclidean space-time $E$
whose fundamental tensor is $\eta_{\alpha \beta }$   ,
$\Gamma _{\alpha \beta }^{\gamma }$    are the Christoffel symbols of the
Riemannian space-time $V$ of dimenstion $n$ , whose  fundamental tensor is
$g_{\alpha \beta }$ . The comma in  eqs. (\ref{myeqs})   denotes  the
covariant differentiation in $E$.

    A peculiarity  of eqs.(\ref{myeqs}) is that they  are invariant under
arbitrary
transformations of the tensor $g_{\alpha \beta }$   retaining invariant  the
equations of
motion of a test particle , i.e.geodesics in $V$ . In  other words,
eqs.(\ref{myeqs})
are the geodesic-invariant . Thus , the tensor field $g_{\alpha \beta }$   is
determined up
to   geodesic  mappings of space-time $V$ (in the way analogous  to  that
the   potential  $A$   in  electrodynamics  is determined up  to  gauge
transformations).
Therefore, additional conditions can be imposed on the tensor
$g_{\alpha \beta }$   . In
partucular, \cite{Verozub} if  the tensor  $g_{\alpha \beta }$   satisfies
the conditions
\begin{equation}
Q_{\alpha } = \Gamma _{\alpha \sigma }^{\sigma } -
 \stackrel{\circ}{\Gamma} _{\alpha \sigma }^{\sigma } ,
\label{gaugeConditions}
\end{equation}
then eqs (\ref{myeqs}) will be  reduced  to vacuum Einstein equations
$R_{\alpha \beta } = 0$ ,
where $R_{\alpha \beta }$   is the Richi tensor.
    As distiguished from the $g_{\alpha \beta }$    or
$\Gamma _{\alpha \beta }^{\gamma }$    the tensor
$B_{\alpha \beta }^{\gamma }$   is invariant
under geodesic mappings  of space-time $V$.   It is in a sense analogous to
strength tensor $F_{\alpha \beta }$ in electrodynamics  which is  invariant
  under the gauge transformations.
     Let us find the spherically symmetric solution of  eqs.(\ref{myeqs})
for a
distant observer . We suppose that the observer's  frame of reference is
an inertial one and space-time is  pseudo-Euclidean.  Choose  a  spherical
coordinate system. Then,  if  a  test  particle  lagrangian  is  invariant
under the  mapping $t \to - t$ , the fundamental metric form of space-time
$V$ can be written as
\begin{equation}
ds^2 = -A (dr)^2 - B [( d \theta )^2 + \sin^2(\theta) (d \varphi )^2 ] +
C (dx^0)^2 ,
\label{ds}
\end{equation}
where  $A$,$B$ and $C$ are the   functions of the radial coordinate $r$ .
    Proceeding from the above- stated, we shall find the functions $A$ ,$B$
 and $C$  as  the  solution  of  the  equations system
\begin{equation}
R_{\alpha \beta } = 0.
\label{Einstein Eqs}
\end{equation}
and
\begin{equation}
Q_{\alpha } = 0
\label{AditionalCondition} ,
\end{equation}
which satisfy the conditions:
\begin{equation}
\lim\limits_{r\to \infty} A = 0 \; \lim\limits_{r\to \infty} B = 0 \;
\lim\limits_{r\to \infty} B = 0 .
\label{limitConditions}
\end{equation}

The equations  $R_{11} = 0$  and  $R_{00} = 0$  can be written as
\begin{equation}
BL_{2} - 2CL_{1} = 0
\label{EinstEq1}
\end{equation}
and
\begin{equation}
B'C' - 2BL_{2} = 0 ,
\label{EinstEq2}
\end{equation}
where
\begin{displaymath}
L_{1} = R_{1212} = B'' /2 - (B')^2 /(4B) - B'A' /(4A) ,
\end{displaymath}
\begin{displaymath}
L_{2} = R_{1010} = -C''/2 + C' (AC)' /(4AC)
\end{displaymath}
and the differentiation of $A$ , $B$ and $C$ with respect to $r$  is denoted
by primes.

     Eqs.(\ref{AditionalCondition})   give
\begin{equation}
B^2AC = r^4    .
\label{BACequal}
\end{equation}
    First,the combination of eqs. (\ref{EinstEq1}) and (\ref{EinstEq2}) gives
\begin{displaymath}
4CL_{1} - B'C' = 0  ,
\end{displaymath}
i.e.
\begin{equation}
2B'' - (B')^2 /B - B' (AC)' / (AC) = 0 .
\label{1}
\end{equation}

Also, taking the logarithm of eq.(\ref{BACequal}) we obtain
\begin{equation}
(AC)' /(AC) = 4/r - 2 B'/B = 0 .
\label{2}
\end{equation}
     Equation (\ref{1}) then becomes
\begin{displaymath}
2B'' + (B')^2 /B - 4 B' /r = 0  ,
\end{displaymath}
or
\begin{equation}
u'/u = 4/r ,
\label{3}
\end{equation}
where     $ u = (B')^2 B$ . By using  (\ref{limitConditions}) we find
\begin{equation}
B = (r^3 +\beta ^3 )^{2/3} ,
\label{Bequal}
\end{equation}
where $\beta $ is a constant.

     Next, from eq. (\ref{EinstEq2}) we find by using
eqs.(\ref{limitConditions})
\begin{equation}
C = 1 - \alpha /B^{1/2} ,
\label{Cequal}
\end{equation}
where $\alpha $ is a constant. The non-relativistic limit should be considered
in order to find this constant . The equations of motion of the  particle
will coincide with the  Newtonian ones if
$\alpha  = 2GM/c^2$ .

     Finally, we can find from  eq. (\ref{BACequal}) the function $A$.
 The constant $\beta$  in eq.(\ref{Bequal}) remains indeterminate. Since
$Q_{\alpha }$  is a vector,
the gauge condition (\ref{AditionalCondition}) is  covariant. Different
constants $\beta$  cause the
differnt solutions of eq.(\ref{myeqs}) in the used coordinate system .
Setting $\beta = \alpha $
we shall obtain the functions $A$ , $B$ and $C$ which have no singularity at
$r \to  0$ :
\begin{equation}
A = (f')^2 (1-\alpha /f)^{-1} ,\; B = f^2 ,\; C = 1 - \alpha /f ,
\label{ABCequal}
\end{equation}
where
\begin{displaymath}
f = ( r^{3} + \alpha ^{3} )^{1/3}
\end{displaymath}
and $ f' = df / dr $ .

     We consider  solution (\ref{ABCequal}) as the basis for the analysis
below.

\section{ Equilibrium configurations .}

     The gravity force in the spherically - symmetric field is given by
\begin{equation}
F^{i} = -m(B_{\beta \gamma }^{i} - B_{\alpha \beta }^{0} \dot{x}^{i})
\dot{x}^{\alpha } \dot{x}^{\beta }
\label{F}
\end{equation}
where $\dot{x}^{i} = dx^{i}/dt$  \cite{Verozub} .(Latin indices run from 1
to 3).
The gravitational force affecting the a test particle  of the mass $m$ in
rest ( $\dot{x^{i }}= 0 $ ) is given by
\begin{equation}
 F=-\frac{GmM}{r^2} \left(1-\alpha/f) \right.,
\label{F(r)}
\end{equation}

where $G$ is the gravitational constant, $\alpha=2GM/c^{2}$, $c$ is the speed
of light, $f=(\alpha^3+r^3)^{1/3}$ . Fig. 1 shows the plot of the function
$F_{1} = - (1/2\  \overline r^{2}) (1 - \alpha /f)$  against the distance
$\overline r=r/ \alpha$.

\setlength{\unitlength}{0.240900pt}
\ifx\plotpoint\undefined\newsavebox{\plotpoint}\fi
\sbox{\plotpoint}{\rule[-0.200pt]{0.400pt}{0.400pt}}%
\special{em:linewidth 0.4pt}%
\begin{picture}(1500,900)(0,0)
\tenrm
\put(264,158){\special{em:moveto}}
\put(264,787){\special{em:lineto}}
\put(264,158){\special{em:moveto}}
\put(284,158){\special{em:lineto}}
\put(1436,158){\special{em:moveto}}
\put(1416,158){\special{em:lineto}}
\put(242,158){\makebox(0,0)[r]{-0.22}}
\put(264,215){\special{em:moveto}}
\put(284,215){\special{em:lineto}}
\put(1436,215){\special{em:moveto}}
\put(1416,215){\special{em:lineto}}
\put(242,215){\makebox(0,0)[r]{-0.2}}
\put(264,272){\special{em:moveto}}
\put(284,272){\special{em:lineto}}
\put(1436,272){\special{em:moveto}}
\put(1416,272){\special{em:lineto}}
\put(242,272){\makebox(0,0)[r]{-0.18}}
\put(264,330){\special{em:moveto}}
\put(284,330){\special{em:lineto}}
\put(1436,330){\special{em:moveto}}
\put(1416,330){\special{em:lineto}}
\put(242,330){\makebox(0,0)[r]{-0.16}}
\put(264,387){\special{em:moveto}}
\put(284,387){\special{em:lineto}}
\put(1436,387){\special{em:moveto}}
\put(1416,387){\special{em:lineto}}
\put(242,387){\makebox(0,0)[r]{-0.14}}
\put(264,444){\special{em:moveto}}
\put(284,444){\special{em:lineto}}
\put(1436,444){\special{em:moveto}}
\put(1416,444){\special{em:lineto}}
\put(242,444){\makebox(0,0)[r]{-0.12}}
\put(264,501){\special{em:moveto}}
\put(284,501){\special{em:lineto}}
\put(1436,501){\special{em:moveto}}
\put(1416,501){\special{em:lineto}}
\put(242,501){\makebox(0,0)[r]{-0.1}}
\put(264,558){\special{em:moveto}}
\put(284,558){\special{em:lineto}}
\put(1436,558){\special{em:moveto}}
\put(1416,558){\special{em:lineto}}
\put(242,558){\makebox(0,0)[r]{-0.08}}
\put(264,615){\special{em:moveto}}
\put(284,615){\special{em:lineto}}
\put(1436,615){\special{em:moveto}}
\put(1416,615){\special{em:lineto}}
\put(242,615){\makebox(0,0)[r]{-0.06}}
\put(264,673){\special{em:moveto}}
\put(284,673){\special{em:lineto}}
\put(1436,673){\special{em:moveto}}
\put(1416,673){\special{em:lineto}}
\put(242,673){\makebox(0,0)[r]{-0.04}}
\put(264,730){\special{em:moveto}}
\put(284,730){\special{em:lineto}}
\put(1436,730){\special{em:moveto}}
\put(1416,730){\special{em:lineto}}
\put(242,730){\makebox(0,0)[r]{-0.02}}
\put(264,787){\special{em:moveto}}
\put(284,787){\special{em:lineto}}
\put(1436,787){\special{em:moveto}}
\put(1416,787){\special{em:lineto}}
\put(242,787){\makebox(0,0)[r]{0}}
\put(264,158){\special{em:moveto}}
\put(264,178){\special{em:lineto}}
\put(264,787){\special{em:moveto}}
\put(264,767){\special{em:lineto}}
\put(264,113){\makebox(0,0){0}}
\put(498,158){\special{em:moveto}}
\put(498,178){\special{em:lineto}}
\put(498,787){\special{em:moveto}}
\put(498,767){\special{em:lineto}}
\put(498,113){\makebox(0,0){2}}
\put(733,158){\special{em:moveto}}
\put(733,178){\special{em:lineto}}
\put(733,787){\special{em:moveto}}
\put(733,767){\special{em:lineto}}
\put(733,113){\makebox(0,0){4}}
\put(967,158){\special{em:moveto}}
\put(967,178){\special{em:lineto}}
\put(967,787){\special{em:moveto}}
\put(967,767){\special{em:lineto}}
\put(967,113){\makebox(0,0){6}}
\put(1202,158){\special{em:moveto}}
\put(1202,178){\special{em:lineto}}
\put(1202,787){\special{em:moveto}}
\put(1202,767){\special{em:lineto}}
\put(1202,113){\makebox(0,0){8}}
\put(1436,158){\special{em:moveto}}
\put(1436,178){\special{em:lineto}}
\put(1436,787){\special{em:moveto}}
\put(1436,767){\special{em:lineto}}
\put(1436,113){\makebox(0,0){10}}
\put(264,158){\special{em:moveto}}
\put(1436,158){\special{em:lineto}}
\put(1436,787){\special{em:lineto}}
\put(264,787){\special{em:lineto}}
\put(264,158){\special{em:lineto}}
\put(850,68){\makebox(0,0){$\overline r$}}
\put(381,730){\makebox(0,0)[r]{$F_{1}$}}
\put(403,730){\special{em:moveto}}
\put(469,730){\special{em:lineto}}
\put(276,692){\special{em:moveto}}
\put(287,597){\special{em:lineto}}
\put(299,506){\special{em:lineto}}
\put(311,421){\special{em:lineto}}
\put(323,347){\special{em:lineto}}
\put(334,286){\special{em:lineto}}
\put(346,241){\special{em:lineto}}
\put(358,212){\special{em:lineto}}
\put(369,198){\special{em:lineto}}
\put(381,197){\special{em:lineto}}
\put(393,206){\special{em:lineto}}
\put(405,222){\special{em:lineto}}
\put(416,244){\special{em:lineto}}
\put(428,268){\special{em:lineto}}
\put(440,293){\special{em:lineto}}
\put(452,319){\special{em:lineto}}
\put(463,345){\special{em:lineto}}
\put(475,370){\special{em:lineto}}
\put(487,393){\special{em:lineto}}
\put(498,416){\special{em:lineto}}
\put(510,437){\special{em:lineto}}
\put(522,457){\special{em:lineto}}
\put(534,475){\special{em:lineto}}
\put(545,493){\special{em:lineto}}
\put(557,509){\special{em:lineto}}
\put(569,524){\special{em:lineto}}
\put(580,538){\special{em:lineto}}
\put(592,551){\special{em:lineto}}
\put(604,563){\special{em:lineto}}
\put(616,574){\special{em:lineto}}
\put(627,584){\special{em:lineto}}
\put(639,594){\special{em:lineto}}
\put(651,603){\special{em:lineto}}
\put(662,612){\special{em:lineto}}
\put(674,620){\special{em:lineto}}
\put(686,627){\special{em:lineto}}
\put(698,634){\special{em:lineto}}
\put(709,641){\special{em:lineto}}
\put(721,647){\special{em:lineto}}
\put(733,653){\special{em:lineto}}
\put(745,658){\special{em:lineto}}
\put(756,663){\special{em:lineto}}
\put(768,668){\special{em:lineto}}
\put(780,673){\special{em:lineto}}
\put(791,677){\special{em:lineto}}
\put(803,681){\special{em:lineto}}
\put(815,685){\special{em:lineto}}
\put(827,689){\special{em:lineto}}
\put(838,692){\special{em:lineto}}
\put(850,695){\special{em:lineto}}
\put(862,699){\special{em:lineto}}
\put(873,702){\special{em:lineto}}
\put(885,704){\special{em:lineto}}
\put(897,707){\special{em:lineto}}
\put(909,710){\special{em:lineto}}
\put(920,712){\special{em:lineto}}
\put(932,714){\special{em:lineto}}
\put(944,717){\special{em:lineto}}
\put(955,719){\special{em:lineto}}
\put(967,721){\special{em:lineto}}
\put(979,723){\special{em:lineto}}
\put(991,725){\special{em:lineto}}
\put(1002,726){\special{em:lineto}}
\put(1014,728){\special{em:lineto}}
\put(1026,730){\special{em:lineto}}
\put(1038,731){\special{em:lineto}}
\put(1049,733){\special{em:lineto}}
\put(1061,734){\special{em:lineto}}
\put(1073,736){\special{em:lineto}}
\put(1084,737){\special{em:lineto}}
\put(1096,738){\special{em:lineto}}
\put(1108,740){\special{em:lineto}}
\put(1120,741){\special{em:lineto}}
\put(1131,742){\special{em:lineto}}
\put(1143,743){\special{em:lineto}}
\put(1155,744){\special{em:lineto}}
\put(1166,745){\special{em:lineto}}
\put(1178,746){\special{em:lineto}}
\put(1190,747){\special{em:lineto}}
\put(1202,748){\special{em:lineto}}
\put(1213,749){\special{em:lineto}}
\put(1225,750){\special{em:lineto}}
\put(1237,750){\special{em:lineto}}
\put(1248,751){\special{em:lineto}}
\put(1260,752){\special{em:lineto}}
\put(1272,753){\special{em:lineto}}
\put(1284,754){\special{em:lineto}}
\put(1295,754){\special{em:lineto}}
\put(1307,755){\special{em:lineto}}
\put(1319,756){\special{em:lineto}}
\put(1331,756){\special{em:lineto}}
\put(1342,757){\special{em:lineto}}
\put(1354,757){\special{em:lineto}}
\put(1366,758){\special{em:lineto}}
\put(1377,759){\special{em:lineto}}
\put(1389,759){\special{em:lineto}}
\put(1401,760){\special{em:lineto}}
\put(1413,760){\special{em:lineto}}
\put(1424,761){\special{em:lineto}}
\put(1436,761){\special{em:lineto}}
\end{picture}

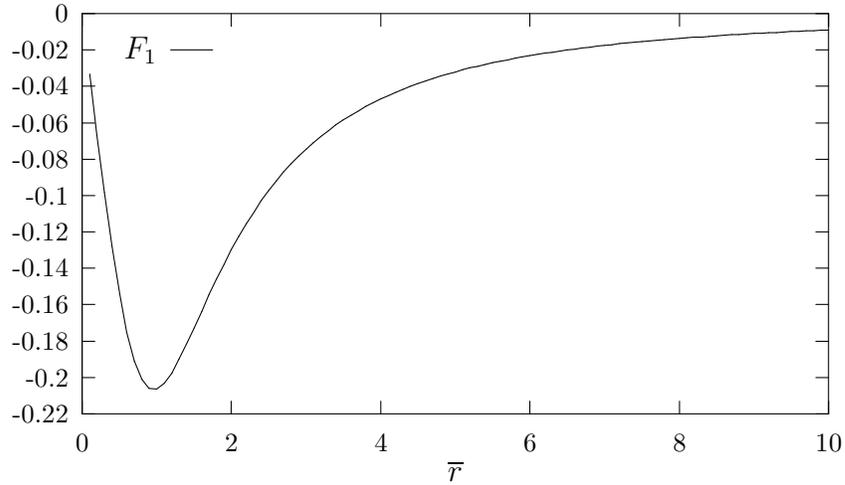
Fig.1  The plot of the function $F_{1}$ against the $\overline r = r/ \alpha$
\vskip 1cm

It follows from Fig.1 that the $|F|$ reaches its maximum at $r$ of the order
of $\alpha$ and tends to zero at $ r \to  0 $.
It would therefore be interesting to khow what the limiting masses of
the equilibrium configurations the gravitational force $F(r)$\ (\ref{F(r)})
can admit.
To answer this question we start from the equation
\begin{equation}
  \frac{dp}{dr} = -\frac{G \rho M}{r^2} \left(1-\alpha/f) \right.
\label{dp/dr=}
\end{equation}
In this equation $\rho$ is the pressure , $M=M(r)$ is the matter mass inside
of a sphere of the radious $r$, $\rho=\rho(r)$ is the matter density at the
distance $r$ from the center, $\alpha$ and $f$ is the function of $M(r)$.

Suppose the equation of state is $p=K\rho^\Gamma$, where  $K$ and $\Gamma$
are constants. For numerical estimates we shall use
their values \cite{Teuk} :

For a degenerated electron gas:

$\Gamma=5/3$ \   $K=1\cdot 10^{13}$ SGS units at $\rho\ll\rho_{0}$, where
$\rho_{0}=10^{6}\   gm/cm^{3} $,

$\Gamma=4/3$ \   $K=1\cdot 10^{15}$ SGS units at  $\rho\gg\rho_{0}$.

For degenerated neutron gas:

$\Gamma=5/3$ \ $K=5\cdot 10^{9}$ SGS units at $\rho\ll\rho_{0}$, where
$\rho_{0}=5\cdot 10^{15}  gm/cm^{3} $ ,

$\Gamma=4/3$ \ $K=1\cdot 10^{15}$ SGS units at $\rho\gg\rho_{0}$.

For rough estimates   we replace $dp/dr$ by $-p/r$, where $p$ is
the average matter pressure and $R$ is its radius. Under the circumstances
we obtain from eq.(\ref{dp/dr=})
\begin{equation}
  \frac{p}{\rho c^{2}} = \frac{\alpha}{2R} \left (1-\alpha/f)\right. .
\label{consequenceEqs2}
\end{equation}

If $ R \gg \alpha $ , then the term $\alpha /f$ is negligible. Setting
      $M(R)\approx \rho R^{3}$ we find the mass of equilibrium states as a
function of $\rho$ :
\begin{equation}
    M=(K/G)^{3/2} \rho ^{ (\Gamma - 4/3)(3/2)}  .
\label{M1=}
\end{equation}

It follows from eq.(\ref{M1=}) that there is the maximal mass
\cite{Chandrasekhar} $M=(K/G)^{3/2}$ at $\rho \gg \rho _{0} $ .

However, according to eq.(\ref{consequenceEqs2}) , there are also equilibrium
configurations  at
$R < \alpha$ . In partucilar, at $R \ll \alpha $ we find from
eq.(\ref{consequenceEqs2}) that
the masses of the equilibrium configurations are given by
\begin{equation}
 M= c^{9/2} 10 ^{-1} K^{-3/4} G^{-3/2} \rho ^{-(\Gamma -1/3)(3/4)} .
\label{M2=}
\end{equation}
These are the configurations with very large masses. For example, the
following equilibrium configurations can be found:

the nonrelativistic electrons: $\rho =10^{5} gm/cm^{3}$,
$M=1.3\cdot 10^{42} gm$, $R=2,3\cdot 10^{12}  cm$,

the relativistic electrons: $\rho = 10^{7} gm/cm^{3}$,
$M=2.3\cdot10^{40} gm$, $R=1.3\cdot 10^{11} cm$,

the nonrelativistic neutrons: $\rho =10^{14} gm/cm^{3}$
$M=3.9\cdot10^{35} gm$, $R=1.6\cdot 10^{7} cm$.

The reason of the two types of configurations existence can be seen from Fig.
2,
where for $\rho =10^{15}\ gm/cm^{3}$  the plots of right-hand and left-hand
sides of Eq.(3) against  the mass $ M$ are given.

\setlength{\unitlength}{0.240900pt}
\ifx\plotpoint\undefined\newsavebox{\plotpoint}\fi
\sbox{\plotpoint}{\rule[-0.200pt]{0.400pt}{0.400pt}}%
\special{em:linewidth 0.4pt}%
\begin{picture}(1500,900)(0,0)
\tenrm
\put(264,158){\special{em:moveto}}
\put(1436,158){\special{em:lineto}}
\put(264,158){\special{em:moveto}}
\put(264,787){\special{em:lineto}}
\put(264,158){\special{em:moveto}}
\put(284,158){\special{em:lineto}}
\put(1436,158){\special{em:moveto}}
\put(1416,158){\special{em:lineto}}
\put(242,158){\makebox(0,0)[r]{0}}
\put(264,248){\special{em:moveto}}
\put(284,248){\special{em:lineto}}
\put(1436,248){\special{em:moveto}}
\put(1416,248){\special{em:lineto}}
\put(242,248){\makebox(0,0)[r]{0.02}}
\put(264,338){\special{em:moveto}}
\put(284,338){\special{em:lineto}}
\put(1436,338){\special{em:moveto}}
\put(1416,338){\special{em:lineto}}
\put(242,338){\makebox(0,0)[r]{0.04}}
\put(264,428){\special{em:moveto}}
\put(284,428){\special{em:lineto}}
\put(1436,428){\special{em:moveto}}
\put(1416,428){\special{em:lineto}}
\put(242,428){\makebox(0,0)[r]{0.06}}
\put(264,517){\special{em:moveto}}
\put(284,517){\special{em:lineto}}
\put(1436,517){\special{em:moveto}}
\put(1416,517){\special{em:lineto}}
\put(242,517){\makebox(0,0)[r]{0.08}}
\put(264,607){\special{em:moveto}}
\put(284,607){\special{em:lineto}}
\put(1436,607){\special{em:moveto}}
\put(1416,607){\special{em:lineto}}
\put(242,607){\makebox(0,0)[r]{0.1}}
\put(264,697){\special{em:moveto}}
\put(284,697){\special{em:lineto}}
\put(1436,697){\special{em:moveto}}
\put(1416,697){\special{em:lineto}}
\put(242,697){\makebox(0,0)[r]{0.12}}
\put(264,787){\special{em:moveto}}
\put(284,787){\special{em:lineto}}
\put(1436,787){\special{em:moveto}}
\put(1416,787){\special{em:lineto}}
\put(242,787){\makebox(0,0)[r]{0.14}}
\put(264,158){\special{em:moveto}}
\put(264,178){\special{em:lineto}}
\put(264,787){\special{em:moveto}}
\put(264,767){\special{em:lineto}}
\put(264,113){\makebox(0,0){0}}
\put(431,158){\special{em:moveto}}
\put(431,178){\special{em:lineto}}
\put(431,787){\special{em:moveto}}
\put(431,767){\special{em:lineto}}
\put(431,113){\makebox(0,0){1e+34}}
\put(599,158){\special{em:moveto}}
\put(599,178){\special{em:lineto}}
\put(599,787){\special{em:moveto}}
\put(599,767){\special{em:lineto}}
\put(599,113){\makebox(0,0){2e+34}}
\put(766,158){\special{em:moveto}}
\put(766,178){\special{em:lineto}}
\put(766,787){\special{em:moveto}}
\put(766,767){\special{em:lineto}}
\put(766,113){\makebox(0,0){3e+34}}
\put(934,158){\special{em:moveto}}
\put(934,178){\special{em:lineto}}
\put(934,787){\special{em:moveto}}
\put(934,767){\special{em:lineto}}
\put(934,113){\makebox(0,0){4e+34}}
\put(1101,158){\special{em:moveto}}
\put(1101,178){\special{em:lineto}}
\put(1101,787){\special{em:moveto}}
\put(1101,767){\special{em:lineto}}
\put(1101,113){\makebox(0,0){5e+34}}
\put(1269,158){\special{em:moveto}}
\put(1269,178){\special{em:lineto}}
\put(1269,787){\special{em:moveto}}
\put(1269,767){\special{em:lineto}}
\put(1269,113){\makebox(0,0){6e+34}}
\put(1436,158){\special{em:moveto}}
\put(1436,178){\special{em:lineto}}
\put(1436,787){\special{em:moveto}}
\put(1436,767){\special{em:lineto}}
\put(1436,113){\makebox(0,0){7e+34}}
\put(264,158){\special{em:moveto}}
\put(1436,158){\special{em:lineto}}
\put(1436,787){\special{em:lineto}}
\put(264,787){\special{em:lineto}}
\put(264,158){\special{em:lineto}}
\put(850,68){\makebox(0,0){$M\ gm$}}
\put(515,338){\makebox(0,0)[l]{$W_{1}(M)$}}
\put(766,517){\makebox(0,0)[l]{$W_{2}(M)$}}
\put(264,372){\special{em:moveto}}
\put(276,372){\special{em:lineto}}
\put(287,372){\special{em:lineto}}
\put(299,372){\special{em:lineto}}
\put(311,372){\special{em:lineto}}
\put(323,372){\special{em:lineto}}
\put(334,372){\special{em:lineto}}
\put(346,372){\special{em:lineto}}
\put(358,372){\special{em:lineto}}
\put(369,372){\special{em:lineto}}
\put(381,372){\special{em:lineto}}
\put(393,372){\special{em:lineto}}
\put(405,372){\special{em:lineto}}
\put(416,372){\special{em:lineto}}
\put(428,372){\special{em:lineto}}
\put(440,372){\special{em:lineto}}
\put(452,372){\special{em:lineto}}
\put(463,372){\special{em:lineto}}
\put(475,372){\special{em:lineto}}
\put(487,372){\special{em:lineto}}
\put(498,372){\special{em:lineto}}
\put(510,372){\special{em:lineto}}
\put(522,372){\special{em:lineto}}
\put(534,372){\special{em:lineto}}
\put(545,372){\special{em:lineto}}
\put(557,372){\special{em:lineto}}
\put(569,372){\special{em:lineto}}
\put(580,372){\special{em:lineto}}
\put(592,372){\special{em:lineto}}
\put(604,372){\special{em:lineto}}
\put(616,372){\special{em:lineto}}
\put(627,372){\special{em:lineto}}
\put(639,372){\special{em:lineto}}
\put(651,372){\special{em:lineto}}
\put(662,372){\special{em:lineto}}
\put(674,372){\special{em:lineto}}
\put(686,372){\special{em:lineto}}
\put(698,372){\special{em:lineto}}
\put(709,372){\special{em:lineto}}
\put(721,372){\special{em:lineto}}
\put(733,372){\special{em:lineto}}
\put(745,372){\special{em:lineto}}
\put(756,372){\special{em:lineto}}
\put(768,372){\special{em:lineto}}
\put(780,372){\special{em:lineto}}
\put(791,372){\special{em:lineto}}
\put(803,372){\special{em:lineto}}
\put(815,372){\special{em:lineto}}
\put(827,372){\special{em:lineto}}
\put(838,372){\special{em:lineto}}
\put(850,372){\special{em:lineto}}
\put(862,372){\special{em:lineto}}
\put(873,372){\special{em:lineto}}
\put(885,372){\special{em:lineto}}
\put(897,372){\special{em:lineto}}
\put(909,372){\special{em:lineto}}
\put(920,372){\special{em:lineto}}
\put(932,372){\special{em:lineto}}
\put(944,372){\special{em:lineto}}
\put(955,372){\special{em:lineto}}
\put(967,372){\special{em:lineto}}
\put(979,372){\special{em:lineto}}
\put(991,372){\special{em:lineto}}
\put(1002,372){\special{em:lineto}}
\put(1014,372){\special{em:lineto}}
\put(1026,372){\special{em:lineto}}
\put(1038,372){\special{em:lineto}}
\put(1049,372){\special{em:lineto}}
\put(1061,372){\special{em:lineto}}
\put(1073,372){\special{em:lineto}}
\put(1084,372){\special{em:lineto}}
\put(1096,372){\special{em:lineto}}
\put(1108,372){\special{em:lineto}}
\put(1120,372){\special{em:lineto}}
\put(1131,372){\special{em:lineto}}
\put(1143,372){\special{em:lineto}}
\put(1155,372){\special{em:lineto}}
\put(1166,372){\special{em:lineto}}
\put(1178,372){\special{em:lineto}}
\put(1190,372){\special{em:lineto}}
\put(1202,372){\special{em:lineto}}
\put(1213,372){\special{em:lineto}}
\put(1225,372){\special{em:lineto}}
\put(1237,372){\special{em:lineto}}
\put(1248,372){\special{em:lineto}}
\put(1260,372){\special{em:lineto}}
\put(1272,372){\special{em:lineto}}
\put(1284,372){\special{em:lineto}}
\put(1295,372){\special{em:lineto}}
\put(1307,372){\special{em:lineto}}
\put(1319,372){\special{em:lineto}}
\put(1331,372){\special{em:lineto}}
\put(1342,372){\special{em:lineto}}
\put(1354,372){\special{em:lineto}}
\put(1366,372){\special{em:lineto}}
\put(1377,372){\special{em:lineto}}
\put(1389,372){\special{em:lineto}}
\put(1401,372){\special{em:lineto}}
\put(1413,372){\special{em:lineto}}
\put(1424,372){\special{em:lineto}}
\put(1436,372){\special{em:lineto}}
\sbox{\plotpoint}{\rule[-0.500pt]{1.000pt}{1.000pt}}%
\special{em:linewidth 1.0pt}%
\put(264,173){\special{em:moveto}}
\put(276,392){\special{em:lineto}}
\put(287,499){\special{em:lineto}}
\put(299,573){\special{em:lineto}}
\put(311,627){\special{em:lineto}}
\put(323,667){\special{em:lineto}}
\put(334,697){\special{em:lineto}}
\put(346,718){\special{em:lineto}}
\put(358,733){\special{em:lineto}}
\put(369,742){\special{em:lineto}}
\put(381,747){\special{em:lineto}}
\put(393,749){\special{em:lineto}}
\put(405,747){\special{em:lineto}}
\put(416,743){\special{em:lineto}}
\put(428,737){\special{em:lineto}}
\put(440,730){\special{em:lineto}}
\put(452,721){\special{em:lineto}}
\put(463,712){\special{em:lineto}}
\put(475,701){\special{em:lineto}}
\put(487,691){\special{em:lineto}}
\put(498,679){\special{em:lineto}}
\put(510,668){\special{em:lineto}}
\put(522,656){\special{em:lineto}}
\put(534,645){\special{em:lineto}}
\put(545,633){\special{em:lineto}}
\put(557,622){\special{em:lineto}}
\put(569,610){\special{em:lineto}}
\put(580,599){\special{em:lineto}}
\put(592,588){\special{em:lineto}}
\put(604,578){\special{em:lineto}}
\put(616,567){\special{em:lineto}}
\put(627,557){\special{em:lineto}}
\put(639,547){\special{em:lineto}}
\put(651,538){\special{em:lineto}}
\put(662,528){\special{em:lineto}}
\put(674,519){\special{em:lineto}}
\put(686,511){\special{em:lineto}}
\put(698,502){\special{em:lineto}}
\put(709,494){\special{em:lineto}}
\put(721,486){\special{em:lineto}}
\put(733,478){\special{em:lineto}}
\put(745,471){\special{em:lineto}}
\put(756,464){\special{em:lineto}}
\put(768,457){\special{em:lineto}}
\put(780,450){\special{em:lineto}}
\put(791,443){\special{em:lineto}}
\put(803,437){\special{em:lineto}}
\put(815,431){\special{em:lineto}}
\put(827,425){\special{em:lineto}}
\put(838,419){\special{em:lineto}}
\put(850,414){\special{em:lineto}}
\put(862,409){\special{em:lineto}}
\put(873,403){\special{em:lineto}}
\put(885,398){\special{em:lineto}}
\put(897,394){\special{em:lineto}}
\put(909,389){\special{em:lineto}}
\put(920,384){\special{em:lineto}}
\put(932,380){\special{em:lineto}}
\put(944,376){\special{em:lineto}}
\put(955,372){\special{em:lineto}}
\put(967,368){\special{em:lineto}}
\put(979,364){\special{em:lineto}}
\put(991,360){\special{em:lineto}}
\put(1002,356){\special{em:lineto}}
\put(1014,353){\special{em:lineto}}
\put(1026,349){\special{em:lineto}}
\put(1038,346){\special{em:lineto}}
\put(1049,343){\special{em:lineto}}
\put(1061,339){\special{em:lineto}}
\put(1073,336){\special{em:lineto}}
\put(1084,333){\special{em:lineto}}
\put(1096,330){\special{em:lineto}}
\put(1108,328){\special{em:lineto}}
\put(1120,325){\special{em:lineto}}
\put(1131,322){\special{em:lineto}}
\put(1143,320){\special{em:lineto}}
\put(1155,317){\special{em:lineto}}
\put(1166,314){\special{em:lineto}}
\put(1178,312){\special{em:lineto}}
\put(1190,310){\special{em:lineto}}
\put(1202,307){\special{em:lineto}}
\put(1213,305){\special{em:lineto}}
\put(1225,303){\special{em:lineto}}
\put(1237,301){\special{em:lineto}}
\put(1248,299){\special{em:lineto}}
\put(1260,297){\special{em:lineto}}
\put(1272,295){\special{em:lineto}}
\put(1284,293){\special{em:lineto}}
\put(1295,291){\special{em:lineto}}
\put(1307,289){\special{em:lineto}}
\put(1319,287){\special{em:lineto}}
\put(1331,286){\special{em:lineto}}
\put(1342,284){\special{em:lineto}}
\put(1354,282){\special{em:lineto}}
\put(1366,281){\special{em:lineto}}
\put(1377,279){\special{em:lineto}}
\put(1389,277){\special{em:lineto}}
\put(1401,276){\special{em:lineto}}
\put(1413,274){\special{em:lineto}}
\put(1424,273){\special{em:lineto}}
\put(1436,271){\special{em:lineto}}
\end{picture}

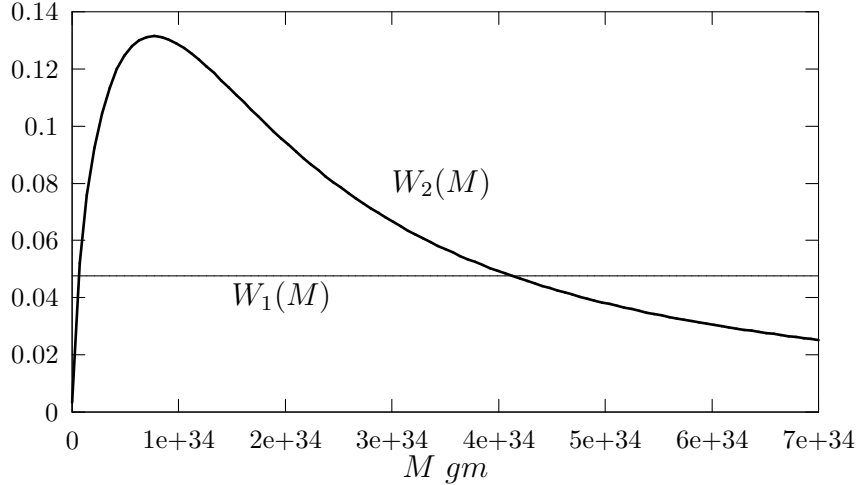
Fig2  The plot of right-hand ($W_{2}(M)$) and left-hand ($W_{1}(M)$) sides of
Eq.(\ref{consequenceEqs2}) against $M$.
\vskip 1cm

The following conclusions can be made after considering the plots of the
above kind:

1. There are no equilibrium configurations whose the density is
larger than a certain value $\rho _{max} \sim 10^{16} gm/cm^{3}$ .

2. For  each value of $\rho < \rho _{max}$ there are two equilibrium
states (with $R > \alpha $  and $R < \alpha $).

Are the configurations with largre masses stable?

The total energy of the degenerate gase is $E=E_{int} + E_{gr}$, where
$E_{int}$ is the intrinsic energy and  $E_{gr}$ is the gravitational
energy. The gravitational energy of a sphere is
\begin{equation}
   E_{gr} = \int_{\infty}^{R} dM(r)\  \chi (r) \ M(r) ,
\label{Egrav}
\end{equation}
where
\begin{displaymath}
 \chi (r) = \int_{\infty}^r dr'\ (r')^{-2} (1 - 1/f) ,
\end{displaymath}
$\alpha  = 2GM(r)/c^{2}$, $f= (\alpha (r) ^3 +(r')^3)^{1/3}$,
\begin{displaymath}
M(r) =4 \pi  \int_{0}^{r} dr' \ \rho \ ( r')^{2} .
\label{M(r)}
\end{displaymath}

The function $\chi (r)$ is approximately

\begin{equation}
   \chi (r) = (1/r)(1-\exp (-r/\alpha )).
\label{chi(r)}
\end{equation}

Therefore, at $p = const$  up to a constant of the order one
\begin{equation}
 E_{gr} = - \frac{G M^{2}}{R}(1 - \exp ( - R/\alpha ).
\label{EgrApprox}
\end{equation}

The intrinsic energy $E_{int} =  \int u\  dM$, where $u$ is the energy per
the mass unit. For the used equation of state
$u = K (\Gamma -1)^{-1} \rho ^{ \Gamma - 1}$ . Thus, up to a constants  of the
order of one
\begin{equation}
E=KM \rho ^{\Gamma -1} - G M^{5/3}\rho ^{1/3}[1-\exp(-QM^{-2/3}\rho ^{-1/3}],
\label{Efinal}
\end{equation}
where $Q = c^2 /2G$ . As an example, Fig.3 and Fig.4 show the plot of the
function
$E = E(\rho )$ for the nonrelativistic neutron gas of the mass
$M=10^{36}\ gm$
and $M = 10^{33}\ gm$ (neutron stares) correspondingly.

\setlength{\unitlength}{0.240900pt}
\ifx\plotpoint\undefined\newsavebox{\plotpoint}\fi
\begin{picture}(1500,900)(0,0)
\tenrm
\put(264,158){\special{em:moveto}}
\put(284,158){\special{em:lineto}}
\put(1436,158){\special{em:moveto}}
\put(1416,158){\special{em:lineto}}
\put(242,158){\makebox(0,0)[r]{-5.98e+56}}
\put(264,248){\special{em:moveto}}
\put(284,248){\special{em:lineto}}
\put(1436,248){\special{em:moveto}}
\put(1416,248){\special{em:lineto}}
\put(242,248){\makebox(0,0)[r]{-5.96e+56}}
\put(264,338){\special{em:moveto}}
\put(284,338){\special{em:lineto}}
\put(1436,338){\special{em:moveto}}
\put(1416,338){\special{em:lineto}}
\put(242,338){\makebox(0,0)[r]{-5.94e+56}}
\put(264,428){\special{em:moveto}}
\put(284,428){\special{em:lineto}}
\put(1436,428){\special{em:moveto}}
\put(1416,428){\special{em:lineto}}
\put(242,428){\makebox(0,0)[r]{-5.92e+56}}
\put(264,517){\special{em:moveto}}
\put(284,517){\special{em:lineto}}
\put(1436,517){\special{em:moveto}}
\put(1416,517){\special{em:lineto}}
\put(242,517){\makebox(0,0)[r]{-5.9e+56}}
\put(264,607){\special{em:moveto}}
\put(284,607){\special{em:lineto}}
\put(1436,607){\special{em:moveto}}
\put(1416,607){\special{em:lineto}}
\put(242,607){\makebox(0,0)[r]{-5.88e+56}}
\put(264,697){\special{em:moveto}}
\put(284,697){\special{em:lineto}}
\put(1436,697){\special{em:moveto}}
\put(1416,697){\special{em:lineto}}
\put(242,697){\makebox(0,0)[r]{-5.86e+56}}
\put(264,787){\special{em:moveto}}
\put(284,787){\special{em:lineto}}
\put(1436,787){\special{em:moveto}}
\put(1416,787){\special{em:lineto}}
\put(242,787){\makebox(0,0)[r]{-5.84e+56}}
\put(264,158){\special{em:moveto}}
\put(264,178){\special{em:lineto}}
\put(264,787){\special{em:moveto}}
\put(264,767){\special{em:lineto}}
\put(264,113){\makebox(0,0){1e+14}}
\put(394,158){\special{em:moveto}}
\put(394,178){\special{em:lineto}}
\put(394,787){\special{em:moveto}}
\put(394,767){\special{em:lineto}}
\put(394,113){\makebox(0,0){2e+14}}
\put(524,158){\special{em:moveto}}
\put(524,178){\special{em:lineto}}
\put(524,787){\special{em:moveto}}
\put(524,767){\special{em:lineto}}
\put(524,113){\makebox(0,0){3e+14}}
\put(655,158){\special{em:moveto}}
\put(655,178){\special{em:lineto}}
\put(655,787){\special{em:moveto}}
\put(655,767){\special{em:lineto}}
\put(655,113){\makebox(0,0){4e+14}}
\put(785,158){\special{em:moveto}}
\put(785,178){\special{em:lineto}}
\put(785,787){\special{em:moveto}}
\put(785,767){\special{em:lineto}}
\put(785,113){\makebox(0,0){5e+14}}
\put(915,158){\special{em:moveto}}
\put(915,178){\special{em:lineto}}
\put(915,787){\special{em:moveto}}
\put(915,767){\special{em:lineto}}
\put(915,113){\makebox(0,0){6e+14}}
\put(1045,158){\special{em:moveto}}
\put(1045,178){\special{em:lineto}}
\put(1045,787){\special{em:moveto}}
\put(1045,767){\special{em:lineto}}
\put(1045,113){\makebox(0,0){7e+14}}
\put(1176,158){\special{em:moveto}}
\put(1176,178){\special{em:lineto}}
\put(1176,787){\special{em:moveto}}
\put(1176,767){\special{em:lineto}}
\put(1176,113){\makebox(0,0){8e+14}}
\put(1306,158){\special{em:moveto}}
\put(1306,178){\special{em:lineto}}
\put(1306,787){\special{em:moveto}}
\put(1306,767){\special{em:lineto}}
\put(1306,113){\makebox(0,0){9e+14}}
\put(1436,158){\special{em:moveto}}
\put(1436,178){\special{em:lineto}}
\put(1436,787){\special{em:moveto}}
\put(1436,767){\special{em:lineto}}
\put(1436,113){\makebox(0,0){1e+15}}
\put(264,158){\special{em:moveto}}
\put(1436,158){\special{em:lineto}}
\put(1436,787){\special{em:lineto}}
\put(264,787){\special{em:lineto}}
\put(264,158){\special{em:lineto}}
\put(850,68){\makebox(0,0){$ \rho\  gm/cm^{3}$}}
\put(524,697){\makebox(0,0)[r]{$E\ erg$}}
\put(546,697){\special{em:moveto}}
\put(612,697){\special{em:lineto}}
\put(264,543){\special{em:moveto}}
\put(276,493){\special{em:lineto}}
\put(287,449){\special{em:lineto}}
\put(299,411){\special{em:lineto}}
\put(311,379){\special{em:lineto}}
\put(323,351){\special{em:lineto}}
\put(334,326){\special{em:lineto}}
\put(346,304){\special{em:lineto}}
\put(358,286){\special{em:lineto}}
\put(369,269){\special{em:lineto}}
\put(381,255){\special{em:lineto}}
\put(393,243){\special{em:lineto}}
\put(405,232){\special{em:lineto}}
\put(416,223){\special{em:lineto}}
\put(428,215){\special{em:lineto}}
\put(440,209){\special{em:lineto}}
\put(452,203){\special{em:lineto}}
\put(463,199){\special{em:lineto}}
\put(475,195){\special{em:lineto}}
\put(487,192){\special{em:lineto}}
\put(498,190){\special{em:lineto}}
\put(510,189){\special{em:lineto}}
\put(522,189){\special{em:lineto}}
\put(534,189){\special{em:lineto}}
\put(545,189){\special{em:lineto}}
\put(557,190){\special{em:lineto}}
\put(569,192){\special{em:lineto}}
\put(580,194){\special{em:lineto}}
\put(592,196){\special{em:lineto}}
\put(604,199){\special{em:lineto}}
\put(616,202){\special{em:lineto}}
\put(627,206){\special{em:lineto}}
\put(639,210){\special{em:lineto}}
\put(651,214){\special{em:lineto}}
\put(662,218){\special{em:lineto}}
\put(674,223){\special{em:lineto}}
\put(686,228){\special{em:lineto}}
\put(698,233){\special{em:lineto}}
\put(709,238){\special{em:lineto}}
\put(721,244){\special{em:lineto}}
\put(733,249){\special{em:lineto}}
\put(745,255){\special{em:lineto}}
\put(756,261){\special{em:lineto}}
\put(768,268){\special{em:lineto}}
\put(780,274){\special{em:lineto}}
\put(791,281){\special{em:lineto}}
\put(803,287){\special{em:lineto}}
\put(815,294){\special{em:lineto}}
\put(827,301){\special{em:lineto}}
\put(838,308){\special{em:lineto}}
\put(850,316){\special{em:lineto}}
\put(862,323){\special{em:lineto}}
\put(873,330){\special{em:lineto}}
\put(885,338){\special{em:lineto}}
\put(897,345){\special{em:lineto}}
\put(909,353){\special{em:lineto}}
\put(920,361){\special{em:lineto}}
\put(932,369){\special{em:lineto}}
\put(944,377){\special{em:lineto}}
\put(955,384){\special{em:lineto}}
\put(967,393){\special{em:lineto}}
\put(979,401){\special{em:lineto}}
\put(991,409){\special{em:lineto}}
\put(1002,417){\special{em:lineto}}
\put(1014,425){\special{em:lineto}}
\put(1026,434){\special{em:lineto}}
\put(1038,442){\special{em:lineto}}
\put(1049,451){\special{em:lineto}}
\put(1061,459){\special{em:lineto}}
\put(1073,468){\special{em:lineto}}
\put(1084,476){\special{em:lineto}}
\put(1096,485){\special{em:lineto}}
\put(1108,494){\special{em:lineto}}
\put(1120,502){\special{em:lineto}}
\put(1131,511){\special{em:lineto}}
\put(1143,520){\special{em:lineto}}
\put(1155,529){\special{em:lineto}}
\put(1166,538){\special{em:lineto}}
\put(1178,546){\special{em:lineto}}
\put(1190,555){\special{em:lineto}}
\put(1202,564){\special{em:lineto}}
\put(1213,573){\special{em:lineto}}
\put(1225,582){\special{em:lineto}}
\put(1237,591){\special{em:lineto}}
\put(1248,600){\special{em:lineto}}
\put(1260,609){\special{em:lineto}}
\put(1272,618){\special{em:lineto}}
\put(1284,627){\special{em:lineto}}
\put(1295,636){\special{em:lineto}}
\put(1307,646){\special{em:lineto}}
\put(1319,655){\special{em:lineto}}
\put(1331,664){\special{em:lineto}}
\put(1342,673){\special{em:lineto}}
\put(1354,682){\special{em:lineto}}
\put(1366,691){\special{em:lineto}}
\put(1377,701){\special{em:lineto}}
\put(1389,710){\special{em:lineto}}
\put(1401,719){\special{em:lineto}}
\put(1413,728){\special{em:lineto}}
\put(1424,738){\special{em:lineto}}
\put(1436,747){\special{em:lineto}}
\end{picture}

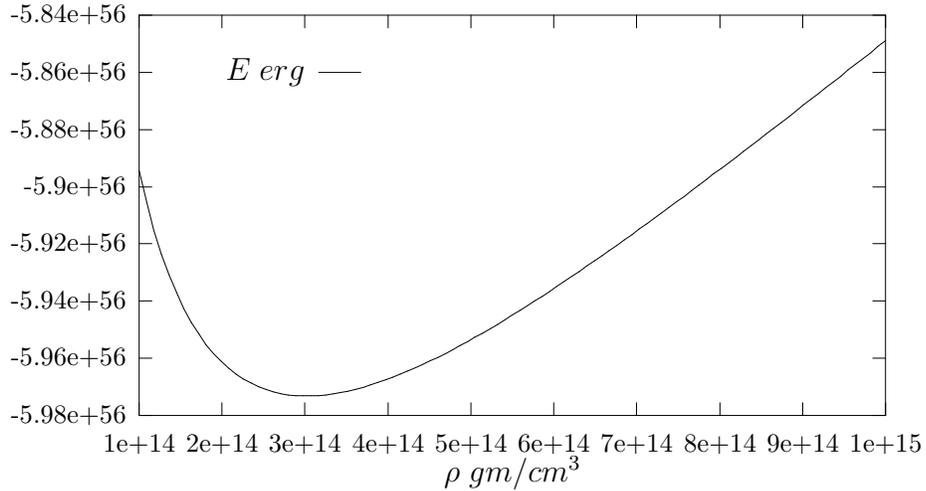
Fig. 3  The plot of the function $E = E(\rho)$ for the neutron configuration
of the mass $M = 10^{36}\ gm$.
\vskip 1cm

\setlength{\unitlength}{0.240900pt}
\ifx\plotpoint\undefined\newsavebox{\plotpoint}\fi
\sbox{\plotpoint}{\rule[-0.200pt]{0.400pt}{0.400pt}}%
\special{em:linewidth 0.4pt}%
\begin{picture}(1500,900)(0,0)
\tenrm
\put(264,158){\special{em:moveto}}
\put(284,158){\special{em:lineto}}
\put(1436,158){\special{em:moveto}}
\put(1416,158){\special{em:lineto}}
\put(242,158){\makebox(0,0)[r]{-2.3e+52}}
\put(264,248){\special{em:moveto}}
\put(284,248){\special{em:lineto}}
\put(1436,248){\special{em:moveto}}
\put(1416,248){\special{em:lineto}}
\put(242,248){\makebox(0,0)[r]{-2.2e+52}}
\put(264,338){\special{em:moveto}}
\put(284,338){\special{em:lineto}}
\put(1436,338){\special{em:moveto}}
\put(1416,338){\special{em:lineto}}
\put(242,338){\makebox(0,0)[r]{-2.1e+52}}
\put(264,428){\special{em:moveto}}
\put(284,428){\special{em:lineto}}
\put(1436,428){\special{em:moveto}}
\put(1416,428){\special{em:lineto}}
\put(242,428){\makebox(0,0)[r]{-2e+52}}
\put(264,517){\special{em:moveto}}
\put(284,517){\special{em:lineto}}
\put(1436,517){\special{em:moveto}}
\put(1416,517){\special{em:lineto}}
\put(242,517){\makebox(0,0)[r]{-1.9e+52}}
\put(264,607){\special{em:moveto}}
\put(284,607){\special{em:lineto}}
\put(1436,607){\special{em:moveto}}
\put(1416,607){\special{em:lineto}}
\put(242,607){\makebox(0,0)[r]{-1.8e+52}}
\put(264,697){\special{em:moveto}}
\put(284,697){\special{em:lineto}}
\put(1436,697){\special{em:moveto}}
\put(1416,697){\special{em:lineto}}
\put(242,697){\makebox(0,0)[r]{-1.7e+52}}
\put(264,787){\special{em:moveto}}
\put(284,787){\special{em:lineto}}
\put(1436,787){\special{em:moveto}}
\put(1416,787){\special{em:lineto}}
\put(242,787){\makebox(0,0)[r]{-1.6e+52}}
\put(264,158){\special{em:moveto}}
\put(264,178){\special{em:lineto}}
\put(264,787){\special{em:moveto}}
\put(264,767){\special{em:lineto}}
\put(264,113){\makebox(0,0){1e+14}}
\put(394,158){\special{em:moveto}}
\put(394,178){\special{em:lineto}}
\put(394,787){\special{em:moveto}}
\put(394,767){\special{em:lineto}}
\put(394,113){\makebox(0,0){2e+14}}
\put(524,158){\special{em:moveto}}
\put(524,178){\special{em:lineto}}
\put(524,787){\special{em:moveto}}
\put(524,767){\special{em:lineto}}
\put(524,113){\makebox(0,0){3e+14}}
\put(655,158){\special{em:moveto}}
\put(655,178){\special{em:lineto}}
\put(655,787){\special{em:moveto}}
\put(655,767){\special{em:lineto}}
\put(655,113){\makebox(0,0){4e+14}}
\put(785,158){\special{em:moveto}}
\put(785,178){\special{em:lineto}}
\put(785,787){\special{em:moveto}}
\put(785,767){\special{em:lineto}}
\put(785,113){\makebox(0,0){5e+14}}
\put(915,158){\special{em:moveto}}
\put(915,178){\special{em:lineto}}
\put(915,787){\special{em:moveto}}
\put(915,767){\special{em:lineto}}
\put(915,113){\makebox(0,0){6e+14}}
\put(1045,158){\special{em:moveto}}
\put(1045,178){\special{em:lineto}}
\put(1045,787){\special{em:moveto}}
\put(1045,767){\special{em:lineto}}
\put(1045,113){\makebox(0,0){7e+14}}
\put(1176,158){\special{em:moveto}}
\put(1176,178){\special{em:lineto}}
\put(1176,787){\special{em:moveto}}
\put(1176,767){\special{em:lineto}}
\put(1176,113){\makebox(0,0){8e+14}}
\put(1306,158){\special{em:moveto}}
\put(1306,178){\special{em:lineto}}
\put(1306,787){\special{em:moveto}}
\put(1306,767){\special{em:lineto}}
\put(1306,113){\makebox(0,0){9e+14}}
\put(1436,158){\special{em:moveto}}
\put(1436,178){\special{em:lineto}}
\put(1436,787){\special{em:moveto}}
\put(1436,767){\special{em:lineto}}
\put(1436,113){\makebox(0,0){1e+15}}
\put(264,158){\special{em:moveto}}
\put(1436,158){\special{em:lineto}}
\put(1436,787){\special{em:lineto}}
\put(264,787){\special{em:lineto}}
\put(264,158){\special{em:lineto}}
\put(850,68){\makebox(0,0){$ \rho\ gm/cm^{3}$}}
\put(524,697){\makebox(0,0)[r]{$E \ erg$}}
\put(546,697){\special{em:moveto}}
\put(612,697){\special{em:lineto}}
\put(264,410){\special{em:moveto}}
\put(276,386){\special{em:lineto}}
\put(287,365){\special{em:lineto}}
\put(299,347){\special{em:lineto}}
\put(311,330){\special{em:lineto}}
\put(323,315){\special{em:lineto}}
\put(334,302){\special{em:lineto}}
\put(346,291){\special{em:lineto}}
\put(358,280){\special{em:lineto}}
\put(369,271){\special{em:lineto}}
\put(381,263){\special{em:lineto}}
\put(393,256){\special{em:lineto}}
\put(405,250){\special{em:lineto}}
\put(416,245){\special{em:lineto}}
\put(428,240){\special{em:lineto}}
\put(440,236){\special{em:lineto}}
\put(452,233){\special{em:lineto}}
\put(463,230){\special{em:lineto}}
\put(475,228){\special{em:lineto}}
\put(487,227){\special{em:lineto}}
\put(498,226){\special{em:lineto}}
\put(510,225){\special{em:lineto}}
\put(522,225){\special{em:lineto}}
\put(534,225){\special{em:lineto}}
\put(545,226){\special{em:lineto}}
\put(557,227){\special{em:lineto}}
\put(569,228){\special{em:lineto}}
\put(580,230){\special{em:lineto}}
\put(592,232){\special{em:lineto}}
\put(604,234){\special{em:lineto}}
\put(616,237){\special{em:lineto}}
\put(627,239){\special{em:lineto}}
\put(639,242){\special{em:lineto}}
\put(651,246){\special{em:lineto}}
\put(662,249){\special{em:lineto}}
\put(674,253){\special{em:lineto}}
\put(686,257){\special{em:lineto}}
\put(698,261){\special{em:lineto}}
\put(709,265){\special{em:lineto}}
\put(721,270){\special{em:lineto}}
\put(733,274){\special{em:lineto}}
\put(745,279){\special{em:lineto}}
\put(756,284){\special{em:lineto}}
\put(768,289){\special{em:lineto}}
\put(780,295){\special{em:lineto}}
\put(791,300){\special{em:lineto}}
\put(803,306){\special{em:lineto}}
\put(815,311){\special{em:lineto}}
\put(827,317){\special{em:lineto}}
\put(838,323){\special{em:lineto}}
\put(850,329){\special{em:lineto}}
\put(862,335){\special{em:lineto}}
\put(873,341){\special{em:lineto}}
\put(885,348){\special{em:lineto}}
\put(897,354){\special{em:lineto}}
\put(909,361){\special{em:lineto}}
\put(920,368){\special{em:lineto}}
\put(932,374){\special{em:lineto}}
\put(944,381){\special{em:lineto}}
\put(955,388){\special{em:lineto}}
\put(967,395){\special{em:lineto}}
\put(979,402){\special{em:lineto}}
\put(991,409){\special{em:lineto}}
\put(1002,417){\special{em:lineto}}
\put(1014,424){\special{em:lineto}}
\put(1026,431){\special{em:lineto}}
\put(1038,439){\special{em:lineto}}
\put(1049,446){\special{em:lineto}}
\put(1061,454){\special{em:lineto}}
\put(1073,462){\special{em:lineto}}
\put(1084,469){\special{em:lineto}}
\put(1096,477){\special{em:lineto}}
\put(1108,485){\special{em:lineto}}
\put(1120,493){\special{em:lineto}}
\put(1131,501){\special{em:lineto}}
\put(1143,509){\special{em:lineto}}
\put(1155,517){\special{em:lineto}}
\put(1166,525){\special{em:lineto}}
\put(1178,533){\special{em:lineto}}
\put(1190,542){\special{em:lineto}}
\put(1202,550){\special{em:lineto}}
\put(1213,558){\special{em:lineto}}
\put(1225,566){\special{em:lineto}}
\put(1237,575){\special{em:lineto}}
\put(1248,583){\special{em:lineto}}
\put(1260,592){\special{em:lineto}}
\put(1272,600){\special{em:lineto}}
\put(1284,609){\special{em:lineto}}
\put(1295,617){\special{em:lineto}}
\put(1307,626){\special{em:lineto}}
\put(1319,635){\special{em:lineto}}
\put(1331,643){\special{em:lineto}}
\put(1342,652){\special{em:lineto}}
\put(1354,661){\special{em:lineto}}
\put(1366,670){\special{em:lineto}}
\put(1377,678){\special{em:lineto}}
\put(1389,687){\special{em:lineto}}
\put(1401,696){\special{em:lineto}}
\put(1413,705){\special{em:lineto}}
\put(1424,714){\special{em:lineto}}
\put(1436,723){\special{em:lineto}}
\end{picture}

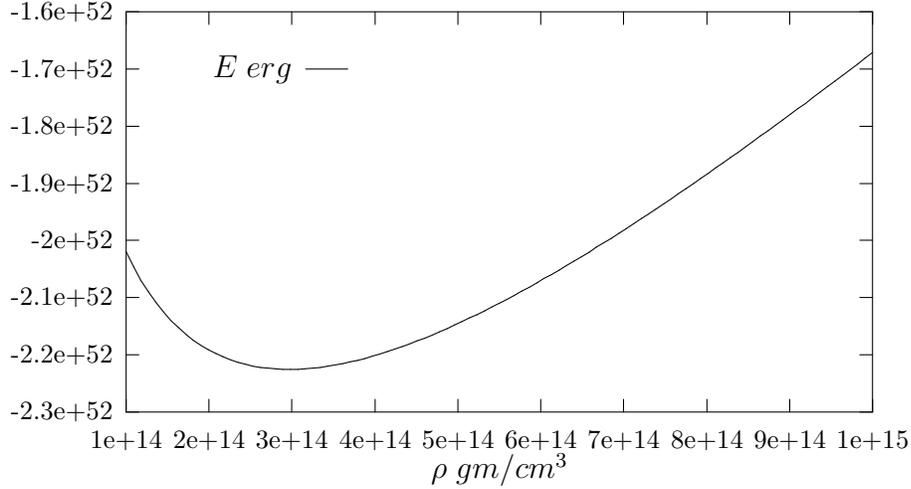
Fig.4  The plot of the function $E = E(\rho)$ for the neutron star of the
mass $M = 10^{33} \ gm$.
\vskip 1 cm

The analysis of  such plots show that the function $E = E (\rho)$ has the
minimum. Thus, the above equilibrium states of large masses  are stable.

\section{Orbits of non-zero  mass    particles.}

     The equations of motion of a test particle of non-zero mass in the
spherically  symmetric field are given by \cite{Verozub}
\begin{equation}
\dot{r}^2 = (c^2 C/A) [ 1 - (C/\overline{E})(1 + \alpha ^2 \overline{J}^2
/B)] ,
\label{EqsMotionTestPart1}
\end{equation}
\begin{equation}
\dot{\varphi } = c\; C \overline{J} \alpha  /(B \overline{E})
\label{EqsMotionTestPart2}
\end{equation}
where $(r,\varphi ,\theta)$ are the spherical coordinates ( $\theta$ is
supposed to be equal to  $\pi/2$) ,
$\dot{r} =dr/dt$ , $\dot{\varphi } = d\varphi /dt$ , $\overline{E} = E/(mc )$
 , $\overline{J} = J /(amc)$ , $E$ is
the particle energy, $J$ is the angular momentum.
     Let $\overline{u} = 1/\overline{f}$ ,  where
$\overline{f} = ( 1 + \overline{r}^3 )^{1/3}$     and $\overline{r} = r/a$ .
Then the differential equation of the  orbits , following from eqs.
(\ref{EqsMotionTestPart1}) and , (\ref{EqsMotionTestPart2})
can be written as
\begin{equation}
d\overline{u} /d \varphi = {\cal G}(\overline{u}) ,
\label{DifEqOrbPart}
\end{equation}
where
\begin{displaymath}
{\cal G}(\overline{u}) = \overline{u}^3 - \overline{u}^2 +
\overline{u} / \overline{J}^2 +
(\overline{E}^2 - 1)/ \overline{J}^2 .
\end{displaymath}

    Eq.(\ref{DifEqOrbPart})  differs from  the  orbit equations of general
relativity
\cite{Chandrasekhar}  by the function $\overline{u}$ instead of the
function
$1/\overline{r}$ . Therefore, the distinction in the orbits becomes apparent
only at the distances  $\overline{r}$ of the order of 1 or less than that.

Setting $\dot{r} = 0$  in eq. (\ref{EqsMotionTestPart1}) we obtain
$\overline{E}^2 =
{\cal N} (\overline{r})$ , where
\begin{equation}
{\cal N }(\overline{r}) = (1 - 1/\overline{f}) (1 + \overline{J}^2 /
\overline{f}^2)
\label{EffPotentialPart}
\end{equation}
is the effective potential \cite{Chandrasekhar} . Fig. 5 shows the
function ${\cal N } = {\cal N }(\overline{r})$ at the typical value of
$\overline{J} = 4$ .

\setlength{\unitlength}{0.240900pt}
\ifx\plotpoint\undefined\newsavebox{\plotpoint}\fi
\sbox{\plotpoint}{\rule[-0.200pt]{0.400pt}{0.400pt}}%
\special{em:linewidth 0.4pt}%
\begin{picture}(1500,900)(0,0)
\tenrm
\put(264,158){\special{em:moveto}}
\put(1436,158){\special{em:lineto}}
\put(264,158){\special{em:moveto}}
\put(264,787){\special{em:lineto}}
\put(264,158){\special{em:moveto}}
\put(284,158){\special{em:lineto}}
\put(1436,158){\special{em:moveto}}
\put(1416,158){\special{em:lineto}}
\put(242,158){\makebox(0,0)[r]{0}}
\put(264,221){\special{em:moveto}}
\put(284,221){\special{em:lineto}}
\put(1436,221){\special{em:moveto}}
\put(1416,221){\special{em:lineto}}
\put(242,221){\makebox(0,0)[r]{0.2}}
\put(264,284){\special{em:moveto}}
\put(284,284){\special{em:lineto}}
\put(1436,284){\special{em:moveto}}
\put(1416,284){\special{em:lineto}}
\put(242,284){\makebox(0,0)[r]{0.4}}
\put(264,347){\special{em:moveto}}
\put(284,347){\special{em:lineto}}
\put(1436,347){\special{em:moveto}}
\put(1416,347){\special{em:lineto}}
\put(242,347){\makebox(0,0)[r]{0.6}}
\put(264,410){\special{em:moveto}}
\put(284,410){\special{em:lineto}}
\put(1436,410){\special{em:moveto}}
\put(1416,410){\special{em:lineto}}
\put(242,410){\makebox(0,0)[r]{0.8}}
\put(264,473){\special{em:moveto}}
\put(284,473){\special{em:lineto}}
\put(1436,473){\special{em:moveto}}
\put(1416,473){\special{em:lineto}}
\put(242,473){\makebox(0,0)[r]{1}}
\put(264,535){\special{em:moveto}}
\put(284,535){\special{em:lineto}}
\put(1436,535){\special{em:moveto}}
\put(1416,535){\special{em:lineto}}
\put(242,535){\makebox(0,0)[r]{1.2}}
\put(264,598){\special{em:moveto}}
\put(284,598){\special{em:lineto}}
\put(1436,598){\special{em:moveto}}
\put(1416,598){\special{em:lineto}}
\put(242,598){\makebox(0,0)[r]{1.4}}
\put(264,661){\special{em:moveto}}
\put(284,661){\special{em:lineto}}
\put(1436,661){\special{em:moveto}}
\put(1416,661){\special{em:lineto}}
\put(242,661){\makebox(0,0)[r]{1.6}}
\put(264,724){\special{em:moveto}}
\put(284,724){\special{em:lineto}}
\put(1436,724){\special{em:moveto}}
\put(1416,724){\special{em:lineto}}
\put(242,724){\makebox(0,0)[r]{1.8}}
\put(264,787){\special{em:moveto}}
\put(284,787){\special{em:lineto}}
\put(1436,787){\special{em:moveto}}
\put(1416,787){\special{em:lineto}}
\put(242,787){\makebox(0,0)[r]{2}}
\put(264,158){\special{em:moveto}}
\put(264,178){\special{em:lineto}}
\put(264,787){\special{em:moveto}}
\put(264,767){\special{em:lineto}}
\put(264,113){\makebox(0,0){0}}
\put(498,158){\special{em:moveto}}
\put(498,178){\special{em:lineto}}
\put(498,787){\special{em:moveto}}
\put(498,767){\special{em:lineto}}
\put(498,113){\makebox(0,0){2}}
\put(733,158){\special{em:moveto}}
\put(733,178){\special{em:lineto}}
\put(733,787){\special{em:moveto}}
\put(733,767){\special{em:lineto}}
\put(733,113){\makebox(0,0){4}}
\put(967,158){\special{em:moveto}}
\put(967,178){\special{em:lineto}}
\put(967,787){\special{em:moveto}}
\put(967,767){\special{em:lineto}}
\put(967,113){\makebox(0,0){6}}
\put(1202,158){\special{em:moveto}}
\put(1202,178){\special{em:lineto}}
\put(1202,787){\special{em:moveto}}
\put(1202,767){\special{em:lineto}}
\put(1202,113){\makebox(0,0){8}}
\put(1436,158){\special{em:moveto}}
\put(1436,178){\special{em:lineto}}
\put(1436,787){\special{em:moveto}}
\put(1436,767){\special{em:lineto}}
\put(1436,113){\makebox(0,0){10}}
\put(264,158){\special{em:moveto}}
\put(1436,158){\special{em:lineto}}
\put(1436,787){\special{em:lineto}}
\put(264,787){\special{em:lineto}}
\put(264,158){\special{em:lineto}}
\put(850,68){\makebox(0,0){$\overline{r}$}}
\put(311,660){\makebox(0,0)[l]{$\overline{E_{2}}$}}
\put(311,734){\makebox(0,0)[l]{$\overline{E_{1}}$}}
\put(850,598){\makebox(0,0)[l]{${\cal N}(\overline{r})$}}
\put(264,158){\special{em:moveto}}
\put(276,182){\special{em:lineto}}
\put(287,225){\special{em:lineto}}
\put(299,279){\special{em:lineto}}
\put(311,340){\special{em:lineto}}
\put(323,403){\special{em:lineto}}
\put(334,464){\special{em:lineto}}
\put(346,520){\special{em:lineto}}
\put(358,567){\special{em:lineto}}
\put(369,605){\special{em:lineto}}
\put(381,633){\special{em:lineto}}
\put(393,654){\special{em:lineto}}
\put(405,667){\special{em:lineto}}
\put(416,674){\special{em:lineto}}
\put(428,677){\special{em:lineto}}
\put(440,676){\special{em:lineto}}
\put(452,673){\special{em:lineto}}
\put(463,668){\special{em:lineto}}
\put(475,663){\special{em:lineto}}
\put(487,656){\special{em:lineto}}
\put(498,649){\special{em:lineto}}
\put(510,642){\special{em:lineto}}
\put(522,635){\special{em:lineto}}
\put(534,628){\special{em:lineto}}
\put(545,621){\special{em:lineto}}
\put(557,614){\special{em:lineto}}
\put(569,607){\special{em:lineto}}
\put(580,601){\special{em:lineto}}
\put(592,595){\special{em:lineto}}
\put(604,589){\special{em:lineto}}
\put(616,584){\special{em:lineto}}
\put(627,579){\special{em:lineto}}
\put(639,574){\special{em:lineto}}
\put(651,569){\special{em:lineto}}
\put(662,565){\special{em:lineto}}
\put(674,561){\special{em:lineto}}
\put(686,557){\special{em:lineto}}
\put(698,553){\special{em:lineto}}
\put(709,549){\special{em:lineto}}
\put(721,546){\special{em:lineto}}
\put(733,543){\special{em:lineto}}
\put(745,539){\special{em:lineto}}
\put(756,537){\special{em:lineto}}
\put(768,534){\special{em:lineto}}
\put(780,531){\special{em:lineto}}
\put(791,529){\special{em:lineto}}
\put(803,526){\special{em:lineto}}
\put(815,524){\special{em:lineto}}
\put(827,522){\special{em:lineto}}
\put(838,520){\special{em:lineto}}
\put(850,518){\special{em:lineto}}
\put(862,516){\special{em:lineto}}
\put(873,514){\special{em:lineto}}
\put(885,513){\special{em:lineto}}
\put(897,511){\special{em:lineto}}
\put(909,510){\special{em:lineto}}
\put(920,508){\special{em:lineto}}
\put(932,507){\special{em:lineto}}
\put(944,505){\special{em:lineto}}
\put(955,504){\special{em:lineto}}
\put(967,503){\special{em:lineto}}
\put(979,502){\special{em:lineto}}
\put(991,501){\special{em:lineto}}
\put(1002,500){\special{em:lineto}}
\put(1014,499){\special{em:lineto}}
\put(1026,498){\special{em:lineto}}
\put(1038,497){\special{em:lineto}}
\put(1049,496){\special{em:lineto}}
\put(1061,495){\special{em:lineto}}
\put(1073,494){\special{em:lineto}}
\put(1084,493){\special{em:lineto}}
\put(1096,493){\special{em:lineto}}
\put(1108,492){\special{em:lineto}}
\put(1120,491){\special{em:lineto}}
\put(1131,490){\special{em:lineto}}
\put(1143,490){\special{em:lineto}}
\put(1155,489){\special{em:lineto}}
\put(1166,489){\special{em:lineto}}
\put(1178,488){\special{em:lineto}}
\put(1190,487){\special{em:lineto}}
\put(1202,487){\special{em:lineto}}
\put(1213,486){\special{em:lineto}}
\put(1225,486){\special{em:lineto}}
\put(1237,485){\special{em:lineto}}
\put(1248,485){\special{em:lineto}}
\put(1260,484){\special{em:lineto}}
\put(1272,484){\special{em:lineto}}
\put(1284,484){\special{em:lineto}}
\put(1295,483){\special{em:lineto}}
\put(1307,483){\special{em:lineto}}
\put(1319,482){\special{em:lineto}}
\put(1331,482){\special{em:lineto}}
\put(1342,482){\special{em:lineto}}
\put(1354,481){\special{em:lineto}}
\put(1366,481){\special{em:lineto}}
\put(1377,481){\special{em:lineto}}
\put(1389,480){\special{em:lineto}}
\put(1401,480){\special{em:lineto}}
\put(1413,480){\special{em:lineto}}
\put(1424,480){\special{em:lineto}}
\put(1436,479){\special{em:lineto}}
\sbox{\plotpoint}{\rule[-0.500pt]{1.000pt}{1.000pt}}%
\special{em:linewidth 1.0pt}%
\put(264,605){\rule{.1pt}{.1pt}}
\put(276,605){\rule{.1pt}{.1pt}}
\put(287,605){\rule{.1pt}{.1pt}}
\put(299,605){\rule{.1pt}{.1pt}}
\put(311,605){\rule{.1pt}{.1pt}}
\put(323,605){\rule{.1pt}{.1pt}}
\put(334,605){\rule{.1pt}{.1pt}}
\put(346,605){\rule{.1pt}{.1pt}}
\put(358,605){\rule{.1pt}{.1pt}}
\put(369,605){\rule{.1pt}{.1pt}}
\sbox{\plotpoint}{\rule[-1.000pt]{2.000pt}{2.000pt}}%
\special{em:linewidth 2.0pt}%
\put(264,693){\rule{.1pt}{.1pt}}
\put(276,693){\rule{.1pt}{.1pt}}
\put(287,693){\rule{.1pt}{.1pt}}
\put(299,693){\rule{.1pt}{.1pt}}
\put(311,693){\rule{.1pt}{.1pt}}
\put(323,693){\rule{.1pt}{.1pt}}
\put(334,693){\rule{.1pt}{.1pt}}
\put(346,693){\rule{.1pt}{.1pt}}
\put(358,693){\rule{.1pt}{.1pt}}
\put(369,693){\rule{.1pt}{.1pt}}
\put(381,693){\rule{.1pt}{.1pt}}
\put(393,693){\rule{.1pt}{.1pt}}
\put(405,693){\rule{.1pt}{.1pt}}
\put(416,693){\rule{.1pt}{.1pt}}
\put(428,693){\rule{.1pt}{.1pt}}
\put(440,693){\rule{.1pt}{.1pt}}
\put(452,693){\rule{.1pt}{.1pt}}
\put(463,693){\rule{.1pt}{.1pt}}
\put(475,693){\rule{.1pt}{.1pt}}
\put(487,693){\rule{.1pt}{.1pt}}
\put(498,693){\rule{.1pt}{.1pt}}
\put(510,693){\rule{.1pt}{.1pt}}
\put(522,693){\rule{.1pt}{.1pt}}
\put(534,693){\rule{.1pt}{.1pt}}
\put(545,693){\rule{.1pt}{.1pt}}
\put(557,693){\rule{.1pt}{.1pt}}
\put(569,693){\rule{.1pt}{.1pt}}
\put(580,693){\rule{.1pt}{.1pt}}
\put(592,693){\rule{.1pt}{.1pt}}
\put(604,693){\rule{.1pt}{.1pt}}
\put(616,693){\rule{.1pt}{.1pt}}
\put(627,693){\rule{.1pt}{.1pt}}
\put(639,693){\rule{.1pt}{.1pt}}
\put(651,693){\rule{.1pt}{.1pt}}
\put(662,693){\rule{.1pt}{.1pt}}
\put(674,693){\rule{.1pt}{.1pt}}
\put(686,693){\rule{.1pt}{.1pt}}
\put(698,693){\rule{.1pt}{.1pt}}
\put(709,693){\rule{.1pt}{.1pt}}
\put(721,693){\rule{.1pt}{.1pt}}
\put(733,693){\rule{.1pt}{.1pt}}
\put(745,693){\rule{.1pt}{.1pt}}
\put(756,693){\rule{.1pt}{.1pt}}
\put(768,693){\rule{.1pt}{.1pt}}
\put(780,693){\rule{.1pt}{.1pt}}
\put(791,693){\rule{.1pt}{.1pt}}
\put(803,693){\rule{.1pt}{.1pt}}
\put(815,693){\rule{.1pt}{.1pt}}
\put(827,693){\rule{.1pt}{.1pt}}
\put(838,693){\rule{.1pt}{.1pt}}
\put(850,693){\rule{.1pt}{.1pt}}
\put(862,693){\rule{.1pt}{.1pt}}
\put(873,693){\rule{.1pt}{.1pt}}
\put(885,693){\rule{.1pt}{.1pt}}
\put(897,693){\rule{.1pt}{.1pt}}
\put(909,693){\rule{.1pt}{.1pt}}
\put(920,693){\rule{.1pt}{.1pt}}
\put(932,693){\rule{.1pt}{.1pt}}
\put(944,693){\rule{.1pt}{.1pt}}
\put(955,693){\rule{.1pt}{.1pt}}
\put(967,693){\rule{.1pt}{.1pt}}
\put(979,693){\rule{.1pt}{.1pt}}
\put(991,693){\rule{.1pt}{.1pt}}
\put(1002,693){\rule{.1pt}{.1pt}}
\put(1014,693){\rule{.1pt}{.1pt}}
\put(1026,693){\rule{.1pt}{.1pt}}
\put(1038,693){\rule{.1pt}{.1pt}}
\put(1049,693){\rule{.1pt}{.1pt}}
\put(1061,693){\rule{.1pt}{.1pt}}
\put(1073,693){\rule{.1pt}{.1pt}}
\put(1084,693){\rule{.1pt}{.1pt}}
\put(1096,693){\rule{.1pt}{.1pt}}
\put(1108,693){\rule{.1pt}{.1pt}}
\put(1120,693){\rule{.1pt}{.1pt}}
\put(1131,693){\rule{.1pt}{.1pt}}
\put(1143,693){\rule{.1pt}{.1pt}}
\put(1155,693){\rule{.1pt}{.1pt}}
\put(1166,693){\rule{.1pt}{.1pt}}
\put(1178,693){\rule{.1pt}{.1pt}}
\put(1190,693){\rule{.1pt}{.1pt}}
\put(1202,693){\rule{.1pt}{.1pt}}
\put(1213,693){\rule{.1pt}{.1pt}}
\put(1225,693){\rule{.1pt}{.1pt}}
\put(1237,693){\rule{.1pt}{.1pt}}
\put(1248,693){\rule{.1pt}{.1pt}}
\put(1260,693){\rule{.1pt}{.1pt}}
\put(1272,693){\rule{.1pt}{.1pt}}
\put(1284,693){\rule{.1pt}{.1pt}}
\put(1295,693){\rule{.1pt}{.1pt}}
\put(1307,693){\rule{.1pt}{.1pt}}
\put(1319,693){\rule{.1pt}{.1pt}}
\put(1331,693){\rule{.1pt}{.1pt}}
\put(1342,693){\rule{.1pt}{.1pt}}
\put(1354,693){\rule{.1pt}{.1pt}}
\put(1366,693){\rule{.1pt}{.1pt}}
\put(1377,693){\rule{.1pt}{.1pt}}
\put(1389,693){\rule{.1pt}{.1pt}}
\put(1401,693){\rule{.1pt}{.1pt}}
\put(1413,693){\rule{.1pt}{.1pt}}
\put(1424,693){\rule{.1pt}{.1pt}}
\put(1436,693){\rule{.1pt}{.1pt}}
\end{picture}

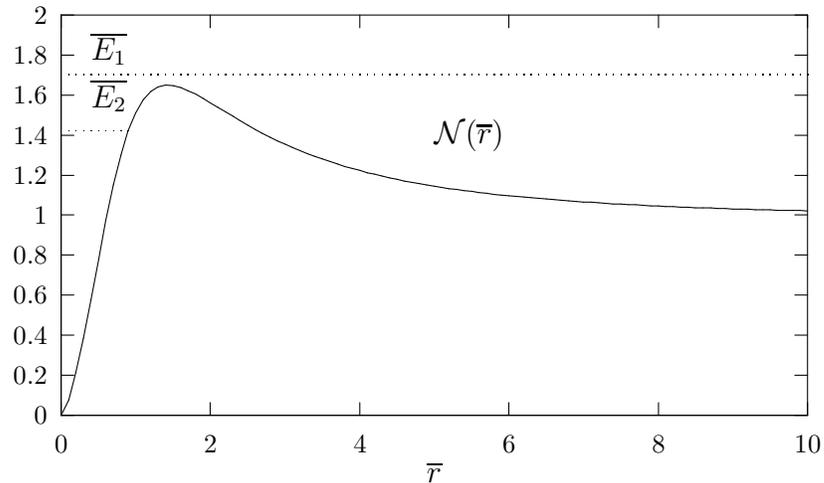
Fig.5  The effective potential ${\cal N} = {\cal N}(\overline{r})$ at
$\overline{J} = 4$. The horizontals $\overline{E} = \overline{E_{1}}$  and
$\overline{E} = \overline{E_{2}}$  show the types of particle orbits due to
lack of the event horizon.
\vskip 1 cm

The function ${\cal N }(\overline{r})$ differs
from the one in general relativity in two respects: 1) it is determined at
every point of the interval $(0, \infty )$ and  2)it tends to zero when
$\overline{r} \to  0 $.

Possible orbit types can be shown by the horizontals $\overline{E} = Const $ .
The
orbits with energies $\overline{E}_{1}$  and $\overline{E}_{2} $  in Fig. 5
have the pecularity due to the
lack of the event horizon. The first orbit type ,with energy
$\overline{E}_{1} $ , exceeding
the maximum  of the function ${\cal \nu }(\overline{r})$ , shows that for each
value of $\overline{J}$ there
exists such value of $\overline{E}$ that the gravitational field cannot  keep
a
particle escaping from the center. Fig.6 shows the difference in the
orbits  of the  above type and the ones in general relativity.

\setlength{\unitlength}{0.240900pt}
\ifx\plotpoint\undefined\newsavebox{\plotpoint}\fi
\sbox{\plotpoint}{\rule[-0.200pt]{0.400pt}{0.400pt}}%
\special{em:linewidth 0.4pt}%
\begin{picture}(1500,900)(0,0)
\tenrm
\put(264,577){\special{em:moveto}}
\put(1436,577){\special{em:lineto}}
\put(655,158){\special{em:moveto}}
\put(655,787){\special{em:lineto}}
\put(264,158){\special{em:moveto}}
\put(284,158){\special{em:lineto}}
\put(1436,158){\special{em:moveto}}
\put(1416,158){\special{em:lineto}}
\put(242,158){\makebox(0,0)[r]{-4}}
\put(264,263){\special{em:moveto}}
\put(284,263){\special{em:lineto}}
\put(1436,263){\special{em:moveto}}
\put(1416,263){\special{em:lineto}}
\put(242,263){\makebox(0,0)[r]{-3}}
\put(264,368){\special{em:moveto}}
\put(284,368){\special{em:lineto}}
\put(1436,368){\special{em:moveto}}
\put(1416,368){\special{em:lineto}}
\put(242,368){\makebox(0,0)[r]{-2}}
\put(264,472){\special{em:moveto}}
\put(284,472){\special{em:lineto}}
\put(1436,472){\special{em:moveto}}
\put(1416,472){\special{em:lineto}}
\put(242,472){\makebox(0,0)[r]{-1}}
\put(264,577){\special{em:moveto}}
\put(284,577){\special{em:lineto}}
\put(1436,577){\special{em:moveto}}
\put(1416,577){\special{em:lineto}}
\put(242,577){\makebox(0,0)[r]{0}}
\put(264,682){\special{em:moveto}}
\put(284,682){\special{em:lineto}}
\put(1436,682){\special{em:moveto}}
\put(1416,682){\special{em:lineto}}
\put(242,682){\makebox(0,0)[r]{1}}
\put(264,787){\special{em:moveto}}
\put(284,787){\special{em:lineto}}
\put(1436,787){\special{em:moveto}}
\put(1416,787){\special{em:lineto}}
\put(242,787){\makebox(0,0)[r]{2}}
\put(264,158){\special{em:moveto}}
\put(264,178){\special{em:lineto}}
\put(264,787){\special{em:moveto}}
\put(264,767){\special{em:lineto}}
\put(264,113){\makebox(0,0){-4}}
\put(459,158){\special{em:moveto}}
\put(459,178){\special{em:lineto}}
\put(459,787){\special{em:moveto}}
\put(459,767){\special{em:lineto}}
\put(459,113){\makebox(0,0){-2}}
\put(655,158){\special{em:moveto}}
\put(655,178){\special{em:lineto}}
\put(655,787){\special{em:moveto}}
\put(655,767){\special{em:lineto}}
\put(655,113){\makebox(0,0){0}}
\put(850,158){\special{em:moveto}}
\put(850,178){\special{em:lineto}}
\put(850,787){\special{em:moveto}}
\put(850,767){\special{em:lineto}}
\put(850,113){\makebox(0,0){2}}
\put(1045,158){\special{em:moveto}}
\put(1045,178){\special{em:lineto}}
\put(1045,787){\special{em:moveto}}
\put(1045,767){\special{em:lineto}}
\put(1045,113){\makebox(0,0){4}}
\put(1241,158){\special{em:moveto}}
\put(1241,178){\special{em:lineto}}
\put(1241,787){\special{em:moveto}}
\put(1241,767){\special{em:lineto}}
\put(1241,113){\makebox(0,0){6}}
\put(1436,158){\special{em:moveto}}
\put(1436,178){\special{em:lineto}}
\put(1436,787){\special{em:moveto}}
\put(1436,767){\special{em:lineto}}
\put(1436,113){\makebox(0,0){8}}
\put(264,158){\special{em:moveto}}
\put(1436,158){\special{em:lineto}}
\put(1436,787){\special{em:lineto}}
\put(264,787){\special{em:lineto}}
\put(264,158){\special{em:lineto}}
\put(850,68){\makebox(0,0){x}}
\put(313,735){\makebox(0,0)[l]{y}}
\put(723,525){\makebox(0,0)[l]{1}}
\put(723,399){\makebox(0,0)[l]{2}}
\put(1436,577){\special{em:moveto}}
\put(1342,489){\special{em:lineto}}
\put(1257,420){\special{em:lineto}}
\put(1181,367){\special{em:lineto}}
\put(1110,325){\special{em:lineto}}
\put(1046,293){\special{em:lineto}}
\put(985,269){\special{em:lineto}}
\put(929,253){\special{em:lineto}}
\put(877,242){\special{em:lineto}}
\put(828,236){\special{em:lineto}}
\put(782,235){\special{em:lineto}}
\put(739,239){\special{em:lineto}}
\put(699,246){\special{em:lineto}}
\put(662,256){\special{em:lineto}}
\put(627,269){\special{em:lineto}}
\put(595,286){\special{em:lineto}}
\put(566,304){\special{em:lineto}}
\put(540,325){\special{em:lineto}}
\put(517,348){\special{em:lineto}}
\put(497,373){\special{em:lineto}}
\put(480,399){\special{em:lineto}}
\put(465,426){\special{em:lineto}}
\put(454,454){\special{em:lineto}}
\put(446,482){\special{em:lineto}}
\put(441,510){\special{em:lineto}}
\put(439,539){\special{em:lineto}}
\put(439,567){\special{em:lineto}}
\put(443,594){\special{em:lineto}}
\put(449,620){\special{em:lineto}}
\put(458,645){\special{em:lineto}}
\put(469,669){\special{em:lineto}}
\put(483,690){\special{em:lineto}}
\put(498,710){\special{em:lineto}}
\put(516,727){\special{em:lineto}}
\put(534,742){\special{em:lineto}}
\put(555,755){\special{em:lineto}}
\put(576,764){\special{em:lineto}}
\put(597,771){\special{em:lineto}}
\put(620,775){\special{em:lineto}}
\put(642,776){\special{em:lineto}}
\put(664,775){\special{em:lineto}}
\put(685,770){\special{em:lineto}}
\put(706,763){\special{em:lineto}}
\put(725,753){\special{em:lineto}}
\put(743,741){\special{em:lineto}}
\put(760,727){\special{em:lineto}}
\put(774,711){\special{em:lineto}}
\put(787,693){\special{em:lineto}}
\put(797,675){\special{em:lineto}}
\put(804,655){\special{em:lineto}}
\put(810,634){\special{em:lineto}}
\put(812,614){\special{em:lineto}}
\put(813,593){\special{em:lineto}}
\put(810,573){\special{em:lineto}}
\put(806,553){\special{em:lineto}}
\put(799,535){\special{em:lineto}}
\put(790,518){\special{em:lineto}}
\put(779,504){\special{em:lineto}}
\put(767,491){\special{em:lineto}}
\put(754,480){\special{em:lineto}}
\put(739,472){\special{em:lineto}}
\put(724,466){\special{em:lineto}}
\put(709,463){\special{em:lineto}}
\put(694,463){\special{em:lineto}}
\put(679,466){\special{em:lineto}}
\put(666,471){\special{em:lineto}}
\put(654,479){\special{em:lineto}}
\put(644,489){\special{em:lineto}}
\put(638,503){\special{em:lineto}}
\put(635,521){\special{em:lineto}}
\put(648,563){\special{em:lineto}}
\sbox{\plotpoint}{\rule[-0.500pt]{1.000pt}{1.000pt}}%
\special{em:linewidth 1.0pt}%
\put(1436,577){\rule{.1pt}{.1pt}}
\put(1342,489){\rule{.1pt}{.1pt}}
\put(1258,420){\rule{.1pt}{.1pt}}
\put(1181,366){\rule{.1pt}{.1pt}}
\put(1111,325){\rule{.1pt}{.1pt}}
\put(1047,292){\rule{.1pt}{.1pt}}
\put(987,268){\rule{.1pt}{.1pt}}
\put(931,251){\rule{.1pt}{.1pt}}
\put(878,240){\rule{.1pt}{.1pt}}
\put(829,234){\rule{.1pt}{.1pt}}
\put(783,233){\rule{.1pt}{.1pt}}
\put(740,236){\rule{.1pt}{.1pt}}
\put(699,242){\rule{.1pt}{.1pt}}
\put(662,252){\rule{.1pt}{.1pt}}
\put(627,265){\rule{.1pt}{.1pt}}
\put(594,282){\rule{.1pt}{.1pt}}
\put(565,300){\rule{.1pt}{.1pt}}
\put(538,321){\rule{.1pt}{.1pt}}
\put(515,344){\rule{.1pt}{.1pt}}
\put(494,369){\rule{.1pt}{.1pt}}
\put(476,395){\rule{.1pt}{.1pt}}
\put(461,422){\rule{.1pt}{.1pt}}
\put(449,451){\rule{.1pt}{.1pt}}
\put(441,479){\rule{.1pt}{.1pt}}
\put(435,509){\rule{.1pt}{.1pt}}
\put(433,538){\rule{.1pt}{.1pt}}
\put(433,566){\rule{.1pt}{.1pt}}
\put(436,594){\rule{.1pt}{.1pt}}
\put(442,622){\rule{.1pt}{.1pt}}
\put(451,648){\rule{.1pt}{.1pt}}
\put(463,672){\rule{.1pt}{.1pt}}
\put(476,694){\rule{.1pt}{.1pt}}
\put(492,715){\rule{.1pt}{.1pt}}
\put(510,733){\rule{.1pt}{.1pt}}
\put(530,749){\rule{.1pt}{.1pt}}
\put(550,762){\rule{.1pt}{.1pt}}
\put(572,772){\rule{.1pt}{.1pt}}
\put(595,780){\rule{.1pt}{.1pt}}
\put(618,784){\rule{.1pt}{.1pt}}
\put(641,785){\rule{.1pt}{.1pt}}
\put(664,784){\rule{.1pt}{.1pt}}
\put(687,779){\rule{.1pt}{.1pt}}
\put(709,772){\rule{.1pt}{.1pt}}
\put(729,762){\rule{.1pt}{.1pt}}
\put(748,750){\rule{.1pt}{.1pt}}
\put(766,735){\rule{.1pt}{.1pt}}
\put(781,719){\rule{.1pt}{.1pt}}
\put(794,700){\rule{.1pt}{.1pt}}
\put(805,681){\rule{.1pt}{.1pt}}
\put(814,660){\rule{.1pt}{.1pt}}
\put(820,638){\rule{.1pt}{.1pt}}
\put(823,616){\rule{.1pt}{.1pt}}
\put(824,594){\rule{.1pt}{.1pt}}
\put(822,572){\rule{.1pt}{.1pt}}
\put(818,552){\rule{.1pt}{.1pt}}
\put(811,532){\rule{.1pt}{.1pt}}
\put(802,513){\rule{.1pt}{.1pt}}
\put(792,496){\rule{.1pt}{.1pt}}
\put(779,482){\rule{.1pt}{.1pt}}
\put(765,469){\rule{.1pt}{.1pt}}
\put(750,459){\rule{.1pt}{.1pt}}
\put(734,451){\rule{.1pt}{.1pt}}
\put(717,445){\rule{.1pt}{.1pt}}
\put(701,443){\rule{.1pt}{.1pt}}
\put(685,442){\rule{.1pt}{.1pt}}
\put(669,445){\rule{.1pt}{.1pt}}
\put(654,449){\rule{.1pt}{.1pt}}
\put(641,456){\rule{.1pt}{.1pt}}
\put(629,464){\rule{.1pt}{.1pt}}
\put(618,473){\rule{.1pt}{.1pt}}
\put(610,484){\rule{.1pt}{.1pt}}
\end{picture}

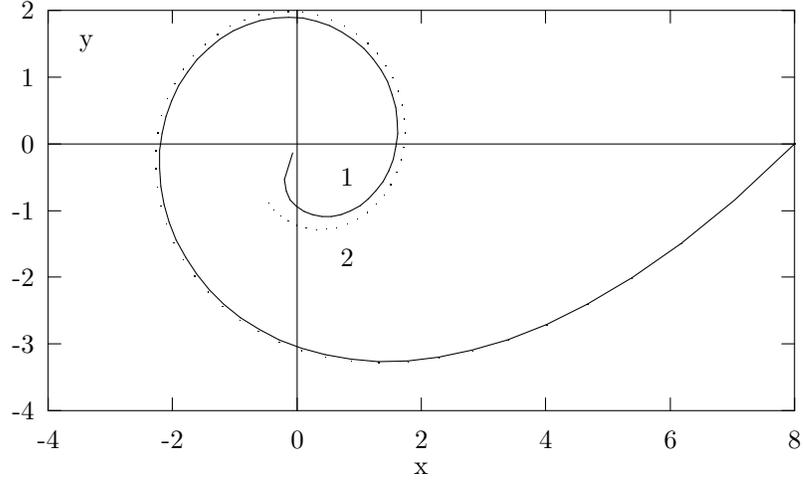
Fig.6   The particle orbits at $\overline{E} = 1$ and $\overline{J} = 1.99$
accordig to eq.(\ref{DifEqOrbPart}) (curve $1$) and the one in general
relativity (curve $2$, dots) . At the distancees $\overline{r} > 1$ the curves
are very close together. The coordinates $x$ and $y$  are determined in the
following way : $x = \overline{r} \cos(\varphi )$ and
$y = \overline{r} \sin(\varphi )$.
\vskip 1 cm

The orbits
of the second type, with the energy $\overline{E}_{2}$  , are the ones of
particls kept  by the  gravitational field near the field center.

    It follows from eqs. (\ref{EqsMotionTestPart1}) and
(\ref{EqsMotionTestPart2}) that the velocity of a test particle freely
falling to the point mass $M$ tends to zero when $r \to 0$ . The time of the
motion of the particle from  some distance $r = r_{0}$  to $r = 0$ is
infinitely large.

    In the  effective potential plot the stable circular orbits are shown
by the minimum points of the function ${\cal N }(\overline{r})$ , the
unstable ones  --- by the
maximum points .The minimums of the function ${\cal N }(\overline{r})$
exists only at
$\overline{J} > \sqrt[3]{3}$   which corresponds to the value of the function
$f(\overline{r}) > 3 $ .
Therefore,  stable circular orbirs exist only at $\overline{r} >
\overline{r}_{cr}$  ,   where
$\overline{r}_{cr} = \sqrt[3]{26} \; \alpha \approx 2.96 \; \alpha $ . The
orbital
 speed of the particle with  $r = r_{cr}$   is equal to $0.4 c$ . At
$r < r_{cr}$   unstable circular orbits can exist. At
$J \to \infty$ the location of the maximums tends to
$\overline{f} = 3/2$ .  Therefore, the
minimum radius of the nonstable circular orbit is $\overline{r}_{min} =
1.33 \alpha $ . (In
general relativity it is equal to $1.5 \alpha $) .The speed of the motion
of a
particle on this orbit is equal to $0.51c$. The binding  energy
$\overline{E} = 0.0572$ , just as it occurs in general relativity.

The rotation frequency  $ \omega  = \dot{\varphi }$ of the circular motion
will be
\begin{equation}
\omega = [ c(\overline{f} - 1) ( \overline{f}^3 \alpha )](\overline{J}/
\overline{E})
\label{omegaParticle}
\end{equation}

      In a circular motion  $\overline{r}$ is the constant and, therefore,
the function
${\cal N } (\overline{r})$ has the minimum. Consequently, from the equation
$d{\cal N} /dr = 0$  we find
\begin{equation}
\overline{J}^2 = \overline{f}^2 / ( 2 \overline{f} - 3)
\label{J}
\end{equation}
Using (\ref{EqsMotionTestPart1}) we have at $\dot{r} = 0$  for
$\overline{E}^2  = {\cal N }$ :
\begin{equation}
\overline{E}^2 = 2 (\overline{f} -1)/ [\overline{f} (2\overline{f} - 1)]
\label{E}
\end{equation}

     From the  eqs. (\ref{EffPotentialPart}) ---(\ref{J}) we obtain
\begin{displaymath}
\omega = \alpha ^{1/2} c / [ \sqrt{2} f^{3/2}].
\end{displaymath}
 Hence, the circular orbits have rotation period
\begin{equation}
T = 2 \sqrt{2} \pi c^{-1} \alpha  (1 + \overline{r}^3)^{1/2}
\label{Tpart}
\end{equation}
( 3 rd Kepler law). In comparison with general relativity the change in $T$
is $2 \% $ at $\overline{r} = 3$  and $20 \% $ at  $\overline{r} = 1.33$ .

     Let us find the apsidal motion. For ellipsoidal orbits the function
${\cal G}(\overline{u})$ has 3 real roots
$\overline{u}_{1} < \overline{u}_{2} < \overline{u}_{3}$
 \cite{Chandrasekhar} .
  The apsidal motion per one period is
\begin{equation}
\delta \varphi = 2 |\Delta | - 2\pi ,
\label{apsMotion}
\end{equation}
where
\begin{equation}
\Delta = \int_{\overline{u}_{1}}^{\overline{u}_{2}} [{\cal G}(\overline{u})]
^{1/2} d \overline{u}
\label{Delta}
\end{equation}

     We have \cite{Byrd}
\begin{equation}
\Delta = 2 (\overline{u}_{3} - \overline{u}_{1})^{1/2} {\cal F}(\pi/2 , q) ,
\label{Delta1}
\end{equation}
where
\begin{equation}
q = [ (\overline{u}_{2} - \overline{u}_{3})/
(\overline{u}_{3} - \overline{u}_{1}) ]^{1/2}
\label{q}
\end{equation}

and
\begin{equation}
{\cal F}( \pi /2,q) = \int_{0}^{\pi/2} (1 - q^{2} \sin^{2}(\beta ))^{1/2}
d \beta
\label{CalF}
\end{equation}

    Let us introduce (by analogy with general relativity) the following
notations:
\begin{equation}
\overline{u}_{1} = (1 - \overline{e })/\overline{p } , \;
\overline{u }_{2} = (1 + \overline{e }/\overline{p } ,   \;
\overline{u }_{3} = 1 - 2 / \overline{p } ,
\label{u1u2u3}
\end{equation}
where the parameter $\overline{p}$ at $\overline{r} \to \infty$ becomes the
focal parameter $p$  divided by $\alpha $
 and parameter $\overline{e}$ becomes the eccentricity $e$  devided by
$\alpha $ .

At  $\overline{r} \gg 1 $  the value of
$q \approx (2 \overline{e} / \overline{p})^{1/2} \ll 1 $ and, therefore,
\begin{equation}
{\cal F}(\pi/2 , q) = (\pi/2)(1+q^2/4+9q^4 /64 + ...)
\label{Fapprox}
\end{equation}

Using   eqs. (\ref{Delta}) and (\ref{CalF}) we find with accuracy up to
$1/\overline{p}^2$
\begin{equation}
\Delta \varphi  = 3 \pi /\overline{p} + (\pi / (8 \overline{p}^2)
(-\overline{e}^2 + 16 \overline{e} + 54 + ... ) .
\label{deltaFi}
\end{equation}

     For the orbits of Mercury or a binary pulsar (such as PSR 1913 + 16 )
the value of $ \overline{u} $  differs very little from the value of
$\alpha /r $ .
Consequently,  the values of
$\overline{p} = 2 / (\overline{u_{1}} + \overline{u_{2}})$   and
$\overline{e} = (\overline{u_{2}} - \overline{u_{1}} /
(\overline{u_{2}} + \overline{u_{1}}) $
differs very little  from the values  of  $ p/ \alpha $  and the
$e/ \alpha $ .
Hence, their  apsidal motion   differs very little  from the general
relativity  prediction. Even, for example, at $\overline{p} = 10$ and
$\overline{e} = 0.5$ , the
difference in  $\Delta \varphi $ is about $6\cdot 10^{-4}  rad$.

\section{ Photon Orbits.}
     The equations of motion of a photon in the spherical symmetric
field are given by \cite{Verozub}
\begin{equation}
\dot{r}^2 = (c^2\;C/A) ( 1-C\;b^2 /B) , \;\;
\dot{\varphi } = c\;C\;b /B ,
\label{eqsMotionPoton}
\end{equation}
where $ b$ is the impact parameter.

     The differential equation of orbits can be written as
\begin{equation}
d \overline{u}/d \varphi  = {\cal G}_{1}(\overline{u}) ,
\label{difEqsMotionPhoton}
\end{equation}
where
\begin{displaymath}
{\cal G}_{1}(\overline{u}) = \overline{u}^3 - \overline{u}^2 +
\overline{b}^2
\end{displaymath}
and $\overline{b} = b / \alpha $.

    Setting $\dot{r} = 0$ in eqs.(\ref{difEqsMotionPhoton}) we obtain
$\overline{b} = (BC)^{1/2} = \overline{f}/ (1 - 1/\overline{f})$. Fig 7
shows $\overline{b}$ as a function of $\overline{r}$ , i.e. the location of
the orbits turning points.

\setlength{\unitlength}{0.240900pt}
\ifx\plotpoint\undefined\newsavebox{\plotpoint}\fi
\sbox{\plotpoint}{\rule[-0.200pt]{0.400pt}{0.400pt}}%
\special{em:linewidth 0.4pt}%
\begin{picture}(1500,900)(0,0)
\tenrm
\put(264,158){\special{em:moveto}}
\put(1436,158){\special{em:lineto}}
\put(264,158){\special{em:moveto}}
\put(264,787){\special{em:lineto}}
\put(264,158){\special{em:moveto}}
\put(284,158){\special{em:lineto}}
\put(1436,158){\special{em:moveto}}
\put(1416,158){\special{em:lineto}}
\put(242,158){\makebox(0,0)[r]{0}}
\put(264,284){\special{em:moveto}}
\put(284,284){\special{em:lineto}}
\put(1436,284){\special{em:moveto}}
\put(1416,284){\special{em:lineto}}
\put(242,284){\makebox(0,0)[r]{2}}
\put(264,410){\special{em:moveto}}
\put(284,410){\special{em:lineto}}
\put(1436,410){\special{em:moveto}}
\put(1416,410){\special{em:lineto}}
\put(242,410){\makebox(0,0)[r]{4}}
\put(264,535){\special{em:moveto}}
\put(284,535){\special{em:lineto}}
\put(1436,535){\special{em:moveto}}
\put(1416,535){\special{em:lineto}}
\put(242,535){\makebox(0,0)[r]{6}}
\put(264,661){\special{em:moveto}}
\put(284,661){\special{em:lineto}}
\put(1436,661){\special{em:moveto}}
\put(1416,661){\special{em:lineto}}
\put(242,661){\makebox(0,0)[r]{8}}
\put(264,787){\special{em:moveto}}
\put(284,787){\special{em:lineto}}
\put(1436,787){\special{em:moveto}}
\put(1416,787){\special{em:lineto}}
\put(242,787){\makebox(0,0)[r]{10}}
\put(264,158){\special{em:moveto}}
\put(264,178){\special{em:lineto}}
\put(264,787){\special{em:moveto}}
\put(264,767){\special{em:lineto}}
\put(264,113){\makebox(0,0){0}}
\put(498,158){\special{em:moveto}}
\put(498,178){\special{em:lineto}}
\put(498,787){\special{em:moveto}}
\put(498,767){\special{em:lineto}}
\put(498,113){\makebox(0,0){2}}
\put(733,158){\special{em:moveto}}
\put(733,178){\special{em:lineto}}
\put(733,787){\special{em:moveto}}
\put(733,767){\special{em:lineto}}
\put(733,113){\makebox(0,0){4}}
\put(967,158){\special{em:moveto}}
\put(967,178){\special{em:lineto}}
\put(967,787){\special{em:moveto}}
\put(967,767){\special{em:lineto}}
\put(967,113){\makebox(0,0){6}}
\put(1202,158){\special{em:moveto}}
\put(1202,178){\special{em:lineto}}
\put(1202,787){\special{em:moveto}}
\put(1202,767){\special{em:lineto}}
\put(1202,113){\makebox(0,0){8}}
\put(1436,158){\special{em:moveto}}
\put(1436,178){\special{em:lineto}}
\put(1436,787){\special{em:moveto}}
\put(1436,767){\special{em:lineto}}
\put(1436,113){\makebox(0,0){10}}
\put(264,158){\special{em:moveto}}
\put(1436,158){\special{em:lineto}}
\put(1436,787){\special{em:lineto}}
\put(264,787){\special{em:lineto}}
\put(264,158){\special{em:lineto}}
\put(850,68){\makebox(0,0){$\overline{r}$}}
\put(381,724){\makebox(0,0)[l]{$b(\overline{r})$}}
\put(381,221){\makebox(0,0)[l]{$b_{1}$ }}
\put(381,410){\makebox(0,0)[l]{$b_{2}$ }}
\put(302,787){\special{em:moveto}}
\put(311,607){\special{em:lineto}}
\put(323,491){\special{em:lineto}}
\put(334,425){\special{em:lineto}}
\put(346,385){\special{em:lineto}}
\put(358,359){\special{em:lineto}}
\put(369,343){\special{em:lineto}}
\put(381,332){\special{em:lineto}}
\put(393,326){\special{em:lineto}}
\put(405,323){\special{em:lineto}}
\put(416,322){\special{em:lineto}}
\put(428,322){\special{em:lineto}}
\put(440,323){\special{em:lineto}}
\put(452,325){\special{em:lineto}}
\put(463,328){\special{em:lineto}}
\put(475,332){\special{em:lineto}}
\put(487,335){\special{em:lineto}}
\put(498,340){\special{em:lineto}}
\put(510,344){\special{em:lineto}}
\put(522,349){\special{em:lineto}}
\put(534,354){\special{em:lineto}}
\put(545,359){\special{em:lineto}}
\put(557,364){\special{em:lineto}}
\put(569,369){\special{em:lineto}}
\put(580,375){\special{em:lineto}}
\put(592,380){\special{em:lineto}}
\put(604,386){\special{em:lineto}}
\put(616,391){\special{em:lineto}}
\put(627,397){\special{em:lineto}}
\put(639,403){\special{em:lineto}}
\put(651,408){\special{em:lineto}}
\put(662,414){\special{em:lineto}}
\put(674,420){\special{em:lineto}}
\put(686,426){\special{em:lineto}}
\put(698,432){\special{em:lineto}}
\put(709,438){\special{em:lineto}}
\put(721,444){\special{em:lineto}}
\put(733,450){\special{em:lineto}}
\put(745,456){\special{em:lineto}}
\put(756,462){\special{em:lineto}}
\put(768,468){\special{em:lineto}}
\put(780,474){\special{em:lineto}}
\put(791,480){\special{em:lineto}}
\put(803,486){\special{em:lineto}}
\put(815,492){\special{em:lineto}}
\put(827,498){\special{em:lineto}}
\put(838,504){\special{em:lineto}}
\put(850,510){\special{em:lineto}}
\put(862,517){\special{em:lineto}}
\put(873,523){\special{em:lineto}}
\put(885,529){\special{em:lineto}}
\put(897,535){\special{em:lineto}}
\put(909,541){\special{em:lineto}}
\put(920,547){\special{em:lineto}}
\put(932,553){\special{em:lineto}}
\put(944,560){\special{em:lineto}}
\put(955,566){\special{em:lineto}}
\put(967,572){\special{em:lineto}}
\put(979,578){\special{em:lineto}}
\put(991,584){\special{em:lineto}}
\put(1002,591){\special{em:lineto}}
\put(1014,597){\special{em:lineto}}
\put(1026,603){\special{em:lineto}}
\put(1038,609){\special{em:lineto}}
\put(1049,615){\special{em:lineto}}
\put(1061,622){\special{em:lineto}}
\put(1073,628){\special{em:lineto}}
\put(1084,634){\special{em:lineto}}
\put(1096,640){\special{em:lineto}}
\put(1108,646){\special{em:lineto}}
\put(1120,653){\special{em:lineto}}
\put(1131,659){\special{em:lineto}}
\put(1143,665){\special{em:lineto}}
\put(1155,671){\special{em:lineto}}
\put(1166,678){\special{em:lineto}}
\put(1178,684){\special{em:lineto}}
\put(1190,690){\special{em:lineto}}
\put(1202,696){\special{em:lineto}}
\put(1213,703){\special{em:lineto}}
\put(1225,709){\special{em:lineto}}
\put(1237,715){\special{em:lineto}}
\put(1248,721){\special{em:lineto}}
\put(1260,727){\special{em:lineto}}
\put(1272,734){\special{em:lineto}}
\put(1284,740){\special{em:lineto}}
\put(1295,746){\special{em:lineto}}
\put(1307,752){\special{em:lineto}}
\put(1319,759){\special{em:lineto}}
\put(1331,765){\special{em:lineto}}
\put(1342,771){\special{em:lineto}}
\put(1354,777){\special{em:lineto}}
\put(1366,784){\special{em:lineto}}
\put(1372,787){\special{em:lineto}}
\sbox{\plotpoint}{\rule[-0.500pt]{1.000pt}{1.000pt}}%
\special{em:linewidth 1.0pt}%
\put(264,284){\rule{.1pt}{.1pt}}
\put(276,284){\rule{.1pt}{.1pt}}
\put(287,284){\rule{.1pt}{.1pt}}
\put(299,284){\rule{.1pt}{.1pt}}
\put(311,284){\rule{.1pt}{.1pt}}
\put(323,284){\rule{.1pt}{.1pt}}
\put(334,284){\rule{.1pt}{.1pt}}
\put(346,284){\rule{.1pt}{.1pt}}
\put(358,284){\rule{.1pt}{.1pt}}
\put(369,284){\rule{.1pt}{.1pt}}
\put(381,284){\rule{.1pt}{.1pt}}
\put(393,284){\rule{.1pt}{.1pt}}
\put(405,284){\rule{.1pt}{.1pt}}
\put(416,284){\rule{.1pt}{.1pt}}
\put(428,284){\rule{.1pt}{.1pt}}
\put(440,284){\rule{.1pt}{.1pt}}
\put(452,284){\rule{.1pt}{.1pt}}
\put(463,284){\rule{.1pt}{.1pt}}
\put(475,284){\rule{.1pt}{.1pt}}
\put(487,284){\rule{.1pt}{.1pt}}
\put(498,284){\rule{.1pt}{.1pt}}
\put(510,284){\rule{.1pt}{.1pt}}
\put(522,284){\rule{.1pt}{.1pt}}
\put(534,284){\rule{.1pt}{.1pt}}
\put(545,284){\rule{.1pt}{.1pt}}
\put(557,284){\rule{.1pt}{.1pt}}
\put(569,284){\rule{.1pt}{.1pt}}
\put(580,284){\rule{.1pt}{.1pt}}
\put(592,284){\rule{.1pt}{.1pt}}
\put(604,284){\rule{.1pt}{.1pt}}
\put(616,284){\rule{.1pt}{.1pt}}
\put(627,284){\rule{.1pt}{.1pt}}
\put(639,284){\rule{.1pt}{.1pt}}
\put(651,284){\rule{.1pt}{.1pt}}
\put(662,284){\rule{.1pt}{.1pt}}
\put(674,284){\rule{.1pt}{.1pt}}
\put(686,284){\rule{.1pt}{.1pt}}
\put(698,284){\rule{.1pt}{.1pt}}
\put(709,284){\rule{.1pt}{.1pt}}
\put(721,284){\rule{.1pt}{.1pt}}
\put(733,284){\rule{.1pt}{.1pt}}
\put(745,284){\rule{.1pt}{.1pt}}
\put(756,284){\rule{.1pt}{.1pt}}
\put(768,284){\rule{.1pt}{.1pt}}
\put(780,284){\rule{.1pt}{.1pt}}
\put(791,284){\rule{.1pt}{.1pt}}
\put(803,284){\rule{.1pt}{.1pt}}
\put(815,284){\rule{.1pt}{.1pt}}
\put(827,284){\rule{.1pt}{.1pt}}
\put(838,284){\rule{.1pt}{.1pt}}
\put(850,284){\rule{.1pt}{.1pt}}
\put(862,284){\rule{.1pt}{.1pt}}
\put(873,284){\rule{.1pt}{.1pt}}
\put(885,284){\rule{.1pt}{.1pt}}
\put(897,284){\rule{.1pt}{.1pt}}
\put(909,284){\rule{.1pt}{.1pt}}
\put(920,284){\rule{.1pt}{.1pt}}
\put(932,284){\rule{.1pt}{.1pt}}
\put(944,284){\rule{.1pt}{.1pt}}
\put(955,284){\rule{.1pt}{.1pt}}
\put(967,284){\rule{.1pt}{.1pt}}
\put(979,284){\rule{.1pt}{.1pt}}
\put(991,284){\rule{.1pt}{.1pt}}
\put(1002,284){\rule{.1pt}{.1pt}}
\put(1014,284){\rule{.1pt}{.1pt}}
\put(1026,284){\rule{.1pt}{.1pt}}
\put(1038,284){\rule{.1pt}{.1pt}}
\put(1049,284){\rule{.1pt}{.1pt}}
\put(1061,284){\rule{.1pt}{.1pt}}
\put(1073,284){\rule{.1pt}{.1pt}}
\put(1084,284){\rule{.1pt}{.1pt}}
\put(1096,284){\rule{.1pt}{.1pt}}
\put(1108,284){\rule{.1pt}{.1pt}}
\put(1120,284){\rule{.1pt}{.1pt}}
\put(1120,284){\rule{.1pt}{.1pt}}
\put(1131,284){\rule{.1pt}{.1pt}}
\put(1143,284){\rule{.1pt}{.1pt}}
\put(1155,284){\rule{.1pt}{.1pt}}
\put(1166,284){\rule{.1pt}{.1pt}}
\put(1178,284){\rule{.1pt}{.1pt}}
\put(1190,284){\rule{.1pt}{.1pt}}
\put(1202,284){\rule{.1pt}{.1pt}}
\put(1213,284){\rule{.1pt}{.1pt}}
\put(1225,284){\rule{.1pt}{.1pt}}
\put(1237,284){\rule{.1pt}{.1pt}}
\put(1248,284){\rule{.1pt}{.1pt}}
\put(1260,284){\rule{.1pt}{.1pt}}
\put(1272,284){\rule{.1pt}{.1pt}}
\put(1284,284){\rule{.1pt}{.1pt}}
\put(1295,284){\rule{.1pt}{.1pt}}
\put(1307,284){\rule{.1pt}{.1pt}}
\put(1319,284){\rule{.1pt}{.1pt}}
\put(1331,284){\rule{.1pt}{.1pt}}
\put(1342,284){\rule{.1pt}{.1pt}}
\put(1354,284){\rule{.1pt}{.1pt}}
\put(1366,284){\rule{.1pt}{.1pt}}
\put(1377,284){\rule{.1pt}{.1pt}}
\put(1389,284){\rule{.1pt}{.1pt}}
\put(1401,284){\rule{.1pt}{.1pt}}
\put(1413,284){\rule{.1pt}{.1pt}}
\put(1319,284){\rule{.1pt}{.1pt}}
\put(1436,284){\rule{.1pt}{.1pt}}
\sbox{\plotpoint}{\rule[-1.000pt]{2.000pt}{2.000pt}}%
\special{em:linewidth 2.0pt}%
\put(264,371){\rule{.1pt}{.1pt}}
\put(276,371){\rule{.1pt}{.1pt}}
\put(287,371){\rule{.1pt}{.1pt}}
\put(299,371){\rule{.1pt}{.1pt}}
\put(311,371){\rule{.1pt}{.1pt}}
\put(323,371){\rule{.1pt}{.1pt}}
\put(334,371){\rule{.1pt}{.1pt}}
\put(346,371){\rule{.1pt}{.1pt}}
\put(352,371){\rule{.1pt}{.1pt}}
\end{picture}

Fig. 7 The function $b = b(\overline{r})$
\vskip 1 cm

The function $\overline{b}(\overline{r})$ , in contrast to  general
relativity, is determined on
all the axes, which is caused by the lack of events horizon.The minimal
value of $\overline{b}$ , i.e. $\overline{b}_{min}$    , is equal to $2.6$ .
It is reached at $\overline{r} = 1.5$ . The
motion of photons can be shown  by the horizontal $\overline{b} = Const$.
In Fig.7
two types of the orbits are shown which have the pecularities due to the
lack of event horizon.The orbit with impact parameter $b = b_{1}$   show that
the attracting mass cannot keep a photon escaping from the center at the
impact parameter $\overline{b} < \overline{b}_{min}$  .This  orbit type also
shows the gravitational
capture of a  photon. The photon finishes at the field center , unlike
general relativity,  where it end on the Schwarzshild sphere.

   The orbit with $b = b_{2}$  shows  that the attracting mass can  keep a
photon escaping from the center only if the impact parameter $\overline{b}$
is greater than $\overline{b}_{min}$     .
      The angle of the light deflection at the distances close to
$\overline{r} = 1.33$  is given by
\begin{equation}
\theta = \ln ( 4.021 /\delta ^2)  ,
\label{theta}
\end{equation}
where \\
$\delta = 2 (\overline{f}_{min} - 3/2)$ and
$\overline{f}_{min} = ( 1 + \overline{r}_{min}^3 )^{1/3}$

\section{ Conclusion Remarks.}
     It follows from eq.(\ref{EgrApprox}) that the gravitational energy
released by a
collapse is more than it was supposed to be hitherto.
     At the same time the gravitational potential  on the surface of a
stable massive configuration of the degenarate Fermi - gas is of the order
of
\begin{equation}
 V = ( G M/R) ( 1 - \exp (-R/\alpha )
\label{V}
\end{equation}
       Hence, it follows from the virial theorem that the above objects
 with  $R \ll \alpha $  are the  ones  with  low temperatures and, therefore,
with
low luminosities. They,more probably, refer  to "dark" matter of the
Universe.
     If their luminosities are caused by an accretion, then the Eddington
limit  of luminosity is approximately
\begin{equation}
{\cal L} = {\cal L}_{Edd}^{0} [ 1 - \exp(-R/\alpha )] ,
\label{luminosity}
\end{equation}
where ${\cal L}_{Edd}^{0} \approx 1\cdot 10^{38} {\cal M}_{\odot} erg/s$ .
Hence, if $R/\alpha \ll 1$ , its luminosity  is
${\cal L } \ll {\cal L}_{Edd}^{0}$     .


\begin{thebibliography}{10}
\bibitem{Rozen} Rozen N. : 1980 , Found. of Phys. 10 , 673
\bibitem{Chang} Chang D.B. and Johnson H.H. : 1980 , Phys. Rev. D , 21 , 874
\bibitem{Verozub} Verozub L.V. :1991 , Phys. Lett.A, 156 , 404
\bibitem{Teuk} Shapiro S.L. and Teukolsky S.A.: 1983, Black Holes, White
Drafts and        Neutron Stars.
\bibitem{Chandrasekhar} Chandrasekhar S : 1983 , The Mathematical Theory of
Black Holes, Oxford        Univ. Press, New York .
\bibitem{Byrd} Byrd P.F. and Fridman M.D.: 1954 , Handbock of elliptic
integrals for  engineers and physicists , Berlin - Gottingen - Heidelberg
\end{thebibliography}
\end{document}